\documentclass[%
 aip,
 amsmath,amssymb,
 reprint,%
]{revtex4-1}
\usepackage{lineno,hyperref}
\usepackage{algorithm}
\usepackage{algorithmicx}
\usepackage{algpseudocode}
\usepackage{color,xcolor}

\usepackage{graphicx}% Include figure files
\usepackage{epstopdf}
\usepackage{dcolumn}% Align table columns on decimal point
\usepackage{bm}% bold math
\usepackage[utf8]{inputenc}
\usepackage[T1]{fontenc}
\usepackage{mathptmx}
\usepackage{lineno,hyperref}
\usepackage{graphics}
\usepackage{amssymb}   % for math
\usepackage{amsmath}
\usepackage[font=scriptsize]{caption}
\usepackage{hyperref}
\usepackage{footnote}
\makesavenoteenv{tabular}
\def\BibTeX{{\rm B\kern-.05em{\sc i\kern-.025em b}\kern-.08emT\kern-.1667em\lower.7ex\hbox{E}\kern-.125emX}}

\begin{document}

\preprint{AIP/123-QED}

\title[A generalized linear threshold model]{A generalized linear threshold model for an improved description of the spreading dynamics}
% Force line breaks with \\
\author{Yijun Ran}%
%\affiliation{College of Computer and Information Science, Southwest University, Beibei, Chongqing, 400715 P. R. China%\\This line break forced with \textbackslash\textbackslash
%}%
\author{Xiaomin Deng}%
%\affiliation{College of Computer and Information Science, Southwest University, Beibei, Chongqing, 400715 P. R. China%\\This line break forced with \textbackslash\textbackslash
%}%
\author{Xiaomeng Wang}
%\altaffiliation{College of Computer and Information Science, Southwest University, Beibei, Chongqing, 400715 P. R. China}%Lines break automatically or can be forced 
\author{Tao Jia}
\email{tjia@swu.edu.cn.}
\affiliation{College of Computer and Information Science, Southwest University, Beibei, Chongqing, 400715 P. R. China%\\This line break forced% with \\
}%

\date{\today}% It is always \today, today,

%
% The abstract is a short summary of the work to be presented in the article.
\begin{abstract}
Many spreading processes in our real-life can be considered as a complex contagion, and the linear threshold (LT) model is often applied as a very representative model for this mechanism. Despite its intensive usage, the LT model suffers several limitations in describing the time evolution of the spreading. First, the discrete-time step that captures the speed of the spreading is vaguely defined. Second, the synchronous updating rule makes the nodes infected in batches, which can not take individual differences into account. Finally, the LT model is incompatible with existing models for the simple contagion. Here we consider a generalized linear threshold (GLT) model for the continuous-time stochastic complex contagion process that can be efficiently implemented by the Gillespie algorithm. The time in this model has a clear mathematical definition and the updating order is rigidly defined. We find that the traditional LT model systematically underestimates the spreading speed and the randomness in the spreading sequence order. We also show that the GLT model works seamlessly with the susceptible-infected (SI) or susceptible-infected-recovered (SIR) model. One can easily combine them to model a hybrid spreading process in which simple contagion accumulates the critical mass for the complex contagion that leads to the global cascades. Overall, the GLT model we proposed can be a useful tool to study complex contagion, especially when studying the time evolution of the spreading. 
\end{abstract}

\maketitle

\begin{quotation}
The linear threshold (LT) model is a typical model for the complex contagion process. However, it systematically underestimates the spreading speed and the randomness in the spreading sequence order. To cope with this issue,  we propose a generalized linear threshold (GLT) model, where the time evolution is controlled by the continuous-time stochastic process. The GLT model can be efficiently implemented by the Gillespie algorithm, providing a useful tool to investigate and simulate more complicated spreading processes, especially when the time evolution is the focus. 
\end{quotation}

\section{\label{sec:level1}Introduction}
The process of adoption such as the adoption of innovations \cite{rogers2010diffusion, weiss2014adoption, zhang2016dynamics}, commercial products \cite{bass1969new, aral2009distinguishing, jin2019emergence} and social behavior \cite{fowler2010cooperative, zheng2013spreading, jia2017quantifying}, and the process of diffusion such as the spread of rumors \cite{ moreno2004dynamics, lazer2018science, vosoughi2018spread}, opinions \cite{travieso2006spread, battiston2020networks} and knowledge \cite{evans2011metaknowledge, iacopini2018network} can all be described as a kind of contagion process \cite{liu2014controlling, guilbeault2018complex, centola2018behavior, centola2010spread, wang2019coevolution}. In these processes, things like information or ideas pass from one person to another through the association between the two individuals, analogous to the infection of diseases. This kind of contagion process is of particular interest when it occurs in sparsely connected networks, where the topology of the network has a big impact on the outcome of the spreading \cite{karsai2011small, lu2011small, xian2019misinformation}, giving rise to a set of interesting phenomena \cite{centola2010spread, castellano2010thresholds}.

The underlying mechanisms generally fall into two categories: simple contagion and complex contagion \cite{borge2013cascading}. The simple contagion is based on disease spreading. An individual, or equivalently a node of a network, has a non-zero probability to be infected if one of the connected neighbors is infected. The infection probability also increases monotonically with the number of infected neighbors. The complex contagion is inspired by collective behaviors in social systems. It assumes that the infection will occur only when some critical mass has reached \cite{watts2002simple}, which can be either the number of contacts \cite{karimi2013threshold, wang2016dynamics} or the number of infected neighbors \cite{granovetter1978threshold}. Correspondingly, the infection probability is non-monotonic, typically captured by a step function that goes directly from 0 to 1 when the critical mass has been reached.

The linear threshold (LT) model is widely used to study complex contagion \cite{granovetter1978threshold, watts2002simple, kempe2003maximizing}. In the model, a node will definitely become infected if the fraction of its infected neighboring nodes goes beyond a threshold value. Previous works using the LT model usually focus on the final consequence of the spreading, such as when the global cascading could occur \cite{watts2002simple, singh2013threshold} or how to select effective seed nodes to maximize the spreading \cite{kempe2003maximizing, liu2018impacts, chen2019identifying}. When it comes to the spreading dynamics, however, the LT model suffers three limitations. First, the evolution in LT model is controlled by discrete-time steps that lack a proper definition, which gives rise to issues when the speed of the spreading needs to be investigated. Second, the status of a node is updated in a synchronous manner. At each time step, all nodes currently satisfying the spreading threshold will turn into the infected state. This can be an issue in application such as machine learning where the order of infection can be important information \cite{cao2017deephawkes}. Finally, the LT model is not very flexible. It is both theoretically and practically challenging if one plans to combine the LT model and other simple contagion model to model some complicated hybrid spreading processes.  

To overcome these limitations, we consider a generalized linear threshold (GLT) model for the continuous-time stochastic spreading process that can be efficiently implemented by the Gillespie algorithm \cite{gillespie1976general, gillespie1977exact}. The evolutionary time in the new model has a physical meaning, which is associated with the rate of the underlying stochastic process. We find that compared with the GLT model, the traditional LT model tends to underestimate the spreading speed. The order of nodes being infected is properly defined in the GLT model, allowing us to better generate synthetic spreading node sequence to model the spreading in real systems. Finally, the GLT model is compatible with the susceptible-infected (SI) or susceptible-infected-recovered (SIR) model \cite{gleeson2011high, gleeson2013binary}, because they are defined under the same mathematical framework. One can easily build a hybrid spreading by combing both simple and complex contagion, or adding the recovery process into the complex contagion. The remainder of the paper is structured as follows. We first give a brief description of the classical LT model with both the synchronous and asynchronous updating rules. We then propose the GLT model and show how to model it efficiently with the Gillespie algorithm. To further shed light on this model, we compare the spreading results from the GLT and LT model. Finally, we show how the GLT model can be combined with other spreading models.  

\section{Results}
\subsection{The linear threshold (LT) model}
The LT model was first introduced in the field of social science \cite{granovetter1978threshold} to analyze the effects of social reinforcement by assuming that each adoption requires a certain fraction of exposures. The community of network science may be more familiar with the work by Duncan Watts \cite{watts2002simple} where the LT model is used to study the condition for global spreading. The same model was also applied in the community of computer science to find the optimal initiator set that maximizes the spreading outcome \cite{kempe2003maximizing}. In the LT model, each node is in one of the two possible states: 0 (inactive, susceptible, \textit{etc.}) or 1 (active, infected, \textit{etc.}). A node $i$ in the network can switch only from state 0 to state 1. The transition probability depends on the fraction of its neighbors that are on state 1, denoted by $\phi_{i}$, as
\begin{equation}
p(\phi_{i})=
\begin{cases}
0&\text{if $\phi_{i} < \phi_{i}^*$} \\
1&\text{if $\phi_{i} \geq \phi_{i}^*$},
\label{equation:probability}
\end{cases}
\end{equation}
where $\phi_{i}^*$ is the threshold value of node $i$, which can be chosen from a probability distribution \cite{wang2019coevolution, watts2002simple} or stay fixed for all nodes. Note that there are other variations for the choice of threshold, such as the number of contacts \cite{karimi2013threshold, wang2016dynamics} or the number of infected neighbors \cite{granovetter1978threshold}. In this paper, we adopt the model by Watts \cite{watts2002simple} that uses the fraction of infected neighbors.

The evolution of the system is characterized by discrete-time steps in the LT model. At each time step, we go through the network and calculate the transition probability of each node according to Eq.(\ref{equation:probability}). All nodes that can change the state are updated synchronously in that time step. The process is repeated until no more nodes can change the state. The time step that characterizes the system evolution, however, is never explicitly defined. This may be because that initial studies \cite{granovetter1978threshold, watts2002simple, kempe2003maximizing} that proposed the model mainly focused on the outcome of the spreading, which does not depend on the choice of time step or how the system actually evolves with time. Nevertheless, the time step needs a proper definition when the spreading dynamics are concerned. Indeed, while the discrete-time step is used in both the LT and SI model, they are inherently different with distinct physical meanings. In the SI model, the time step is associated with the probability that the disease is transferred from one node to another, or the chance that a node gets infected from an infected neighbor. If the time step is equivalent to a longer period of real time, the infection probability would be tuned larger, which eventually gives the same spreading dynamics. As an example, the infection probability in disease spreading would be different if the time step refers to an hour or a day. In the LT model, however, the time step is associated with a node's status updating, which is independent of the transmission probability. Its physical meaning, related to why every node updates its status within one time step, is not clearly interpreted.

Another issue is the order of the infection. Under the synchronous updating rule, all nodes satisfying the threshold condition change the state together in one time step. Considering the case that node $i$ changes the state from 0 to 1, which makes its neighboring node $j$ reach the threshold. While the transition condition is satisfied, node $j$ can not change the state in that time step. In other words, node $j$'s state is frozen till all nodes in the same batch of node $i$ complete the transition. In terms of the infection order, node $j$ always ranks behind them. Note that the infection order is important in tasks such as tracking the spreading source \cite{shen2016locating} or learning the embedding of the underlying network \cite{bourigault2016representation, gou2018learning}. The simplification of LT model may limit its application in generating the synthetic spreading node sequence in real systems. A simple fix of this issue is to use the asynchronous updating rule. One option is that at each time step, we randomly pick only one node from those whose threshold is reached and update the node's state. In this way, the spreading order would be more realistic. However, the spreading dynamics would become unrealistic as the number of infected nodes increases linearly with time steps. An alternative option is to randomly pick an arbitrary node at each time step regardless of its threshold condition and update its states according to its $p(\phi_{i})$. This actually becomes a Monte Carlo simulation \cite{gleeson2013binary, porter2016dynamical}. But the computational complexity raised to $O(N^2)$ where $N$ is the number of nodes in a network. More importantly, even though the asynchronous updating rule can fix the order, it is very difficult to model a system with individual differences. For example, if we assume that some nodes are more active and would change the states faster than others, it would be very difficult to implement this feature in the model. 

Finally, the LT model is not very flexible. This is partially related with the vague definition of the discrete-time step. If we want to model a system with both simple and complex contagion, we need to define two types of time steps. One type of time step is for the deterministic infection in the LT model and the other for the probabilistic infection in the SI model. The conversion between the two types of time step can be an interesting interplay, which, however, lacks a proper definition and brings challenges for theoretical interpretation. Because the node status is updated synchronous at the end of LT time step, the spreading curve will not be smooth but containing multiple bursts separated by fixed time intervals. We also need to propose a rule to decide which action should occur first when the two types of time step coincides. All these difficulties increase when more dynamics are involved, such as adding a recovery process to have the susceptible?infected-susceptible (SIS) or SIR model in the system. Therefore, it is challenging to apply the traditional LT model for complex spreading process.

\subsection{A generalized linear threshold model and its stochastic simulation}
To cope with the issues mentioned, we consider a simple variation of the original LT model and generalize it to continuous-time stochastic process. In the generalized linear threshold (GLT) model, a node $i$ has a certain rate to transfer from state 0 to state 1, which is given by
\begin{equation}
\beta_{i} (\phi_{i})=
\begin{cases}
0& \text{if $\phi_{i} < \phi_{i}^*$}\\
k_i& \text{if $\phi_{i} \geq \phi_{i}^*$}.
\label{equation:rate}
\end{cases}
\end{equation}
Eq.(\ref{equation:rate}) is similar to Eq.(\ref{equation:probability}), both capturing a threshold dynamic. When $\phi_{i}$ is below the threshold, the transition (or infection) can not occur. When $\phi_{i}$ is above the threshold, the transition will occur in certain. The extra information given by Eq.(\ref{equation:rate}) is the rate $k_i$, which controls the speed of the transition and to what extent node $i$ would be infected ahead of other nodes. By assigning different $k_i$ value to different nodes, the individual differences on the transition are well characterized.

To efficiently simulate the GLT model, we apply the Gillespie algorithm \cite{gillespie1976general, gillespie1977exact}. It is an efficient simulation method for the stochastic process and was heavily used to investigate the interactions of molecules in chemical systems \cite{sinitsyn2009adiabatic, ramaswamy2011partial} or the cellular growth and division in biological systems \cite{jia2011intrinsic, qiu2019quantifying, kumar2016frequency}. It can also be used to simulate the epidemic spreading such as SI and SIR model \cite{gleeson2011high, fennell2016limitations, gleeson2013binary}. Indeed, though it is not explicitly specified, when we use a rate $k$ to quantify a dynamic process, we imply that it is a Poisson process with a rate $k$. The inter-event time or waiting time $\tau$ is random and follows an exponential distribution with a rate $k$. This property can be generalized to cases when multiple Poisson processes coexist. Assume that there are $N$ nodes in the network, each has a transition rate $\beta_{i}$. The inter-event time $\tau$ for the occurrence of next transition follows an exponential distribution with rate $\tilde\beta = \sum_{i=1}^N \beta_{i}$. In practice, $\tau$ can be efficiently calculated from a random number $r$ uniformed picked from the interval (0,1) as
\begin{equation}
\tau =  -\frac{\ln{r}}{\tilde\beta}.
\end{equation}
The probability that the transition takes place on node $j$ linearly depends on its transition rate as 
\begin{equation}
p = \frac{\beta_{j}}{\tilde\beta}. 
\label{eq:p}
\end{equation}

\begin{algorithm}[H]
\caption{The generalized linear threshold model based on the Gillespie algorithm}
\label{code:modelcalc}
\begin{algorithmic}[1]
\Require Network $G$, infection rate $\beta$, threshold $\phi^*$, initial seeds $\rho$
\Ensure Time series list T, susceptible number list S, infected number list I
\Function {\textbf{GLT}} {$G$, $\beta$, $\phi^*$, $\rho$}
   \State $T, S, I \gets [0], [\left| G \right| - \rho], [\rho]$ where the $\left| G \right|$ is the number of nodes
   \State $nodes \gets $nodes in the G
   \State $infected\_nodes \gets $random.sample($nodes, \rho$)
   \State $risk\_nodes \gets $the susceptible neighbors of the $infected\_nodes$
   \State $susceptible\_nodes \gets []$
   \For{$u$ \textbf{in} $nodes$}
       \State $infected\_rate[u]  \gets \beta$
   \EndFor
   \State $\tau \gets 0$
   \For{$n$ \textbf{in} $risk\_nodes$}
       \State $num[n] \gets $ the number of infected neighbors of $n$
       \State $degree[n] \gets $ the number of neighbors of $n$
       \If {$\frac{num[n]}{degree[n]} \geq \phi^*$ }
              \State add $n$ into $susceptible\_nodes$
              \State remove $n$ from $risk\_nodes$
       \EndIf
   \EndFor
   \State $total\_rate \gets \sum_{n \in susceptible\_nodes} infected\_rate[n]$  
   \While{ $total\_rate > 0$}
       \State $n = random.choice(susceptible\_nodes)$ \Comment{If each node has different rate $\beta_{i}$ in a network, please see below for an optimization.}
       \State remove $n$ from $susceptible\_nodes$
       \State add $n$ into $infected\_nodes$
       \State $\tau  \gets \tau - \frac{ln (random.uniform(0.0,1.0))}{total\_rate}$
       \State Update $T$, $S$, $I$
       \State $susceptible\_neighbors \gets $the susceptible neighbors of the $n$
       \For{$u$ \textbf{in} $susceptible\_neighbors$}
        		 \If {$u$ \textbf{not in} $susceptible\_nodes$}
		 	\State $risk\_nodes \gets u$
		 \EndIf
       \EndFor 
       \For{$n$ \textbf{in} $risk\_nodes$}
            \State $num[n] \gets $ the number of infected neighbors of $n$
            \State $degree[n] \gets $ the number of neighbors of $n$
            \If {$\frac{num[n]}{degree[n]} \geq \phi^*$ }
                \State add $n$ into $susceptible\_nodes$
                \State remove $n$ from $risk\_nodes$
            \EndIf
       \EndFor
       \State $total\_rate \gets \sum_{n \in susceptible\_nodes} infected\_rate[n]$  
   \EndWhile   
\State \Return  $T$, $S$, $I$
\EndFunction
\end{algorithmic}
\end{algorithm}

The Gillespie algorithm takes this property of the stochastic process. At each simulation step, it decides, in a random manner, which event would occur and when it would occur. The procedure can be summarized as follows:\\
1. At the time $t$, find all events that may occur (with a positive rate) and get the sum of the rate $\tilde\beta$.\\
2. Generate a random variable $\tau$ from an exponential distribution with rate $\tilde\beta$.\\
3. Randomly draw an event according to the probability of $p$ in Eq.(\ref{eq:p})\\
4. Update the system according to the event drawn. Update the time from $t$ to $t+\tau$.\\
5. Repeat from step 1.\\

To illustrate the simulation of the GLT model, we provide the pseudocode in Algorithm \ref{code:modelcalc}. At each simulation step, we need to determine which action would occur from the rates of all actions. When the $k_i$ is the same for all nodes, or there are only a few choice of $k_i$ values, we can do a random selection of actions to simplify this process, which takes only $O(1)$ complexity. In comparison with the Monte Carlo version of the LT model with complexity $O(N^2)$\cite{gleeson2013binary, porter2016dynamical}, the Gillespie algorithm significantly reduces the computation cost.
When all nodes have different $k_i$ values, the Monte Carlo version of the LT model would fail because it assumes that all nodes are picked to update the states with equal probability \cite{gleeson2013binary, porter2016dynamical}. The Gillespie algorithm can handle this situation by deciding the process that happens according the rate $k_i$. The selection is a typical fitness proportionate selection, also known as roulette-wheel selection \cite{lipowski2012roulette}. The complexity is usually $O(N)$ because we need to calculate $\beta_{j}/\tilde\beta$ for every node at each simulation step. However, using a recently proposed optimization, the complexity can be reduced to $O(1)$ type \cite{lipowski2012roulette}. Taken together with the $N$ nodes in the system, the complexity to simulate the whole evolution is roughly $O(N)$.

\subsection{The application of the GLT model}
\begin{figure}[!htb]
\centering
\includegraphics[width=1.0\linewidth]{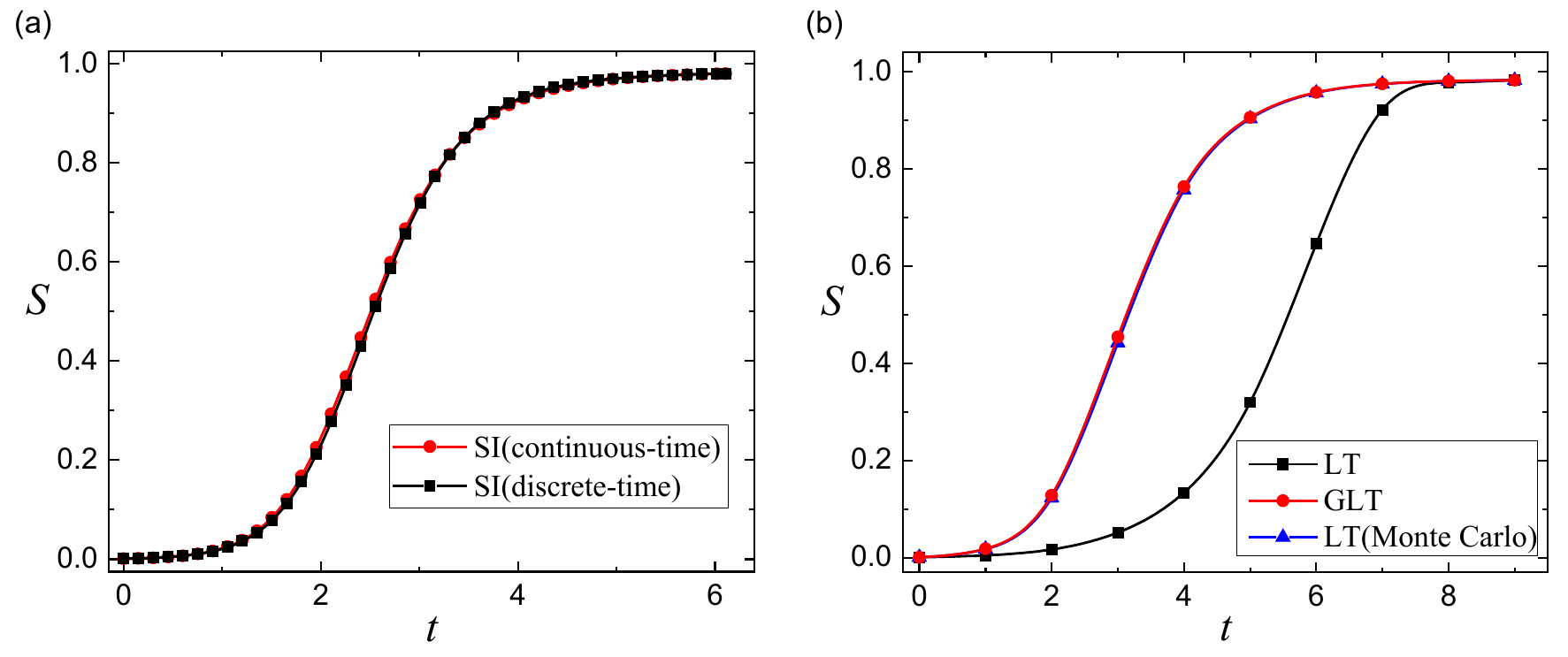}
\caption{Time evolution of the average cascade size $S$ (usually known as infected fraction in epidemic researches) on Erd{\"o}s-R{\'e}nyi network with the size $N=1000$ and the average degree $\langle k \rangle=4$. (a) The continuous-time and discrete-time SI model. The continuous-time SI model is implemented by the Gillespie algorithm in which each node has the same infection rate $\beta = 1$. In the discrete-time SI model, the infection probability is $p = 0.01$ and nodes' states are updated synchronously. (b) The GLT and LT model in which each node has the same threshold value $\phi^*=0.16$. In the GLT model, each node has the same rate $\beta = 1$. All curves in (a) and (b) are based on the average over $10^3$ runs of simulation. In each run, we choose 1 same node as initiator.}
\label{fig:comparing}
\end{figure}

To show features of the GLT model, we compare its time evolution with that of the LT model. Because the transition rate $k_i$ can be any value, we have to first adjust the continuous rate and the discrete-time step to make the continuous-time and discrete-time model comparable. Unlike SI model where the relationship among the rate, the infection probability and the discrete-time step is known \cite{fennell2016limitations}, there is no method yet to handle the parameter conversion in the threshold model. To cope with this issue, we consider re-scaling the spreading time window.
We choose average cascade size $S=0.98$ as our reference point and record the time (either discrete-time steps or continuous time) takes from the beginning of the spreading to $S=0.98$ as the time window. The discrete-time steps and the continuous time are then re-scaled such that the spreading time window is the same. We consider $S=0.98$ instead of $S=1$ to avoid possible tails in the infection of the last node. To test the validity of our approach, we first apply it to SI model. After re-scaling the time window, the continuous-time and discrete-time spreading curve overlap each other (Fig. \ref{fig:comparing}(a)), giving rise to the same results by parameter conversion. Hence the approach of time window re-scaling is effective. For the threshold model, we consider its Monte Carlo version \cite{gleeson2013binary, porter2016dynamical} as the baseline, where an arbitrary node is picked at random at each time step regardless of its threshold condition. The state of the node is then updated according to Eq.(\ref{equation:probability}). This baseline is compared with the spreading generated by the GLT model and the traditional LT model with synchronous updating rule. The dynamics by the GLT model matches with the baseline, but the dynamics of the LT model is different, where the infected size grows slower than both the GLT model and the baseline (Fig. \ref{fig:comparing}(b)). This indicates that the LT model underestimates the speed of spreading, supporting our initial statement of the LT model's limitations.

The LT model underestimates the spreading dynamics due to its coarse-grained description of the time evolution. The spreading speed is not instantaneously updated according to the number of nodes satisfying the threshold condition. As an example, let us assume there are 10 nodes satisfying the threshold in the LT model. Naturally, these 10 nodes will be infected in the next time step. If we assume one time step corresponds to a continuous time $T$, the infection of each of the 10 nodes takes $T/10$ time on average. In the GLT model, if there are 10 nodes satisfying the threshold, the infection rate of the first node will be $10 \times k$ (assuming $k_i = k$ for all nodes). The average waiting time to infect the first node is $T/10$ if $k$ is set as $k=1/T$, which is the same as that in the LT model. However, the infection of one node will activate more nodes during the spreading. Therefore, after the infection of this node, there will be more than 10 nodes in the system satisfying the threshold. The rate for the next infection to occur is greater than $10 \times k$ and the average waiting time takes less than $T/10$ time. As the spreading proceeds, the nodes available for state change accumulates, which presents a higher spreading speed than what LT model suggests.

\begin{figure*}
\centering
\includegraphics[width=16cm]{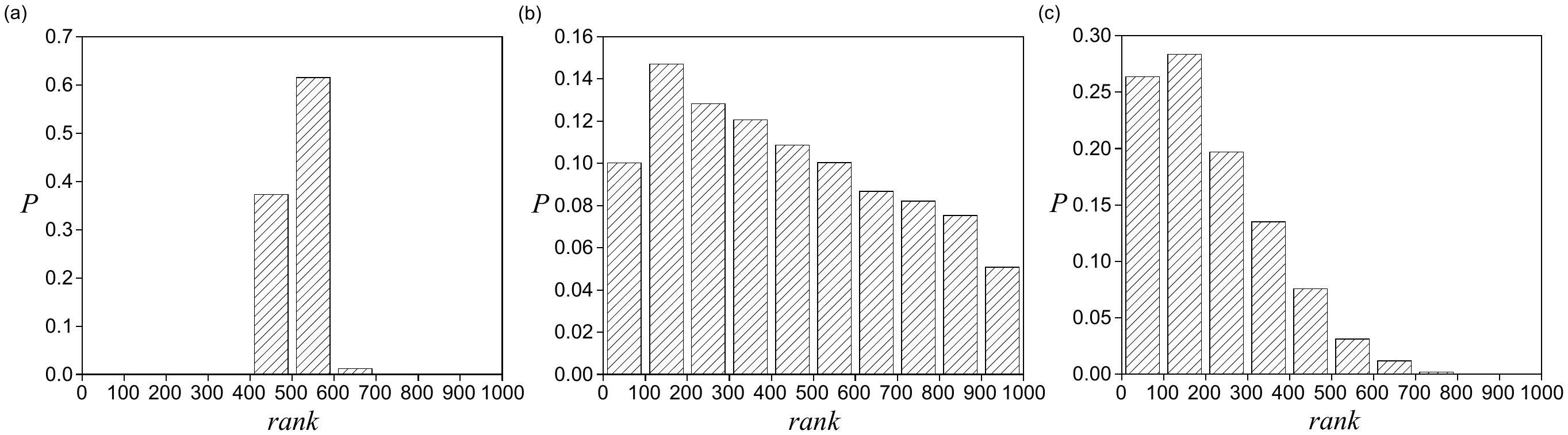}
\caption{The distribution of the rank of an infected node in spreading sequences on an Erd{\"o}s-R{\'e}nyi network with $N=1000$ nodes and the average degree $\langle k \rangle=4$. The threshold value is $\phi^*=0.16$. We fix the node in all models and check when it will be infected. (a) The LT model, (b) the GLT model in which every node has the same infection rate $\beta = 1$, (c) the GLT model in which the node selected has the infection rate $\beta=10$ and other nodes have infection rate $\beta=1$. The results are based on $10^4$ runs of simulation. We select 1 node as the initiator and fix it in each run.}
\label{fig:order}
\end{figure*}

\begin{figure}
\centering
\includegraphics[width=1.0\linewidth]{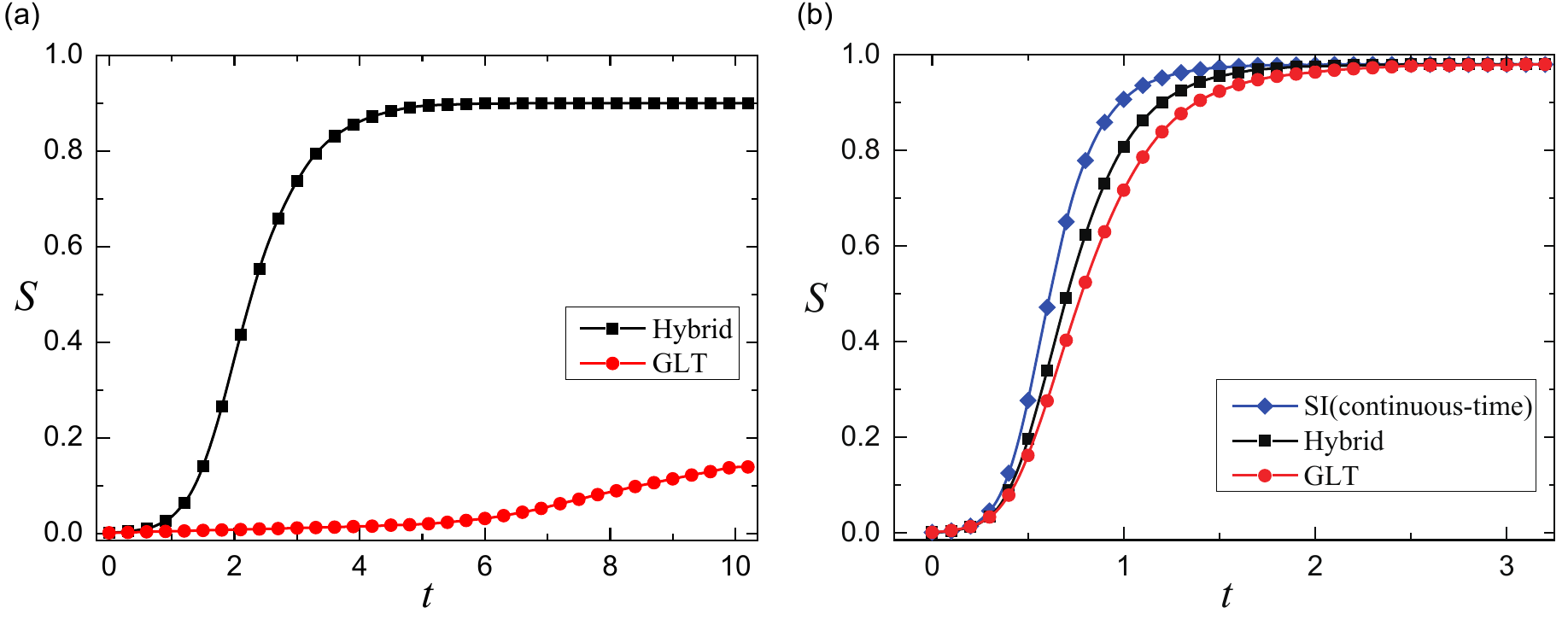}
\caption{Time evolution of the average cascade size $S$ for among the continuous-time SI model (the blue diamond), the hybrid model (the black square) and the GLT model (the red circle) on the Erd{\"o}s-R{\'e}nyi network with $N=1000$ nodes and the average degree $\langle k \rangle=4$. (a) The threshold value $\phi^*=0.25$ and every node has the same infection rate $\beta = 2$ in the GLT and hybrid model. (b) The threshold value $\phi^*=0.16$ and every node has the same infection rate $\beta = 4$. The fraction of nodes that follow the GLT model and the SI model is 1:1 in the hybrid model. We select 1 node as the initiator and fix it in each run. The curve is based on $10^3$ runs of simulation.}
\label{fig:hybrid}
\end{figure}

We also compare the spreading sequence generated by the two models. We fix all other parameters, pick a node, and focus on the rank of the node in the full spreading sequence. Due to the randomness in the spreading, the rank is not the same in every spreading. Hence we measure the distribution of the rank. Note that in the LT model, multiple nodes change the state in a batch. So we randomize the order of infection in the same batch in the LT model. We find that the rank of a node's infection can be very segregated in the LT model, and can be very random in the GLT model (Fig. \ref{fig:order}). Indeed, even when we increase the $k_i$ of the node selected, giving it a much higher priority to change the state, its rank is still very randomly distributed (Fig. \ref{fig:order}(c)). Therefore, when using the LT model to generate synthetic spreading data, we may underestimate the complexity of spreading brought by the inherent randomness. 

Finally, because the GLT model is based on the continuous-time stochastic process, it is compatible with other continuous-time stochastic processes. We only need to add more reactions in the queue when multiple processes coexist. As an example, we apply the GLT model as a tool to simulate the hybrid spreading process. There are works in epidemics assuming that the disease infection rate can be different under different conditions \cite{altizer2006seasonality, freeman2006herpes}. Hence there will be two infection rates in the system. This feature is hard to implement in the LT model but can be easily added by the GLT model. Here we consider a more interesting situation. In the threshold spreading, to have a global cascade triggered by a single initiator, the threshold value $\phi^*$ needs to be small ($\phi^* \le 1/\langle k \rangle$) \cite{wang2019coevolution, watts2002simple}. This brings questions on how a social spreading, which is usually believed to be the complex contagion, could occur since the threshold of a real social system may not be that small. One explanation is that there can be multiple initiators \cite{singh2013threshold, jankowski2018probing, jankowski2017balancing}. Alternatively, we may also assume that both simple and complex contagion are active \cite{liu2018interactive}. Here we analyze co-evolutionary contagion which is a hybrid model combining the SI and GLT model. In the model, we assume that there are two types of nodes, one evolves according to the SI model and the other to the GLT model. The result shows that complex contagion alone can not take place, but the global cascade could occur when simple contagion co-exist (Fig. \ref{fig:hybrid}(a)). The simulation result demonstrates the model's capability in combining other spreading mechanisms.

We further find that the dynamic of the hybrid model always locates between the SI model and the GLT model whatever the infected rate $\beta$ is when the threshold $\phi^*$ is smaller than the $1/\langle k \rangle$ (Fig. \ref{fig:hybrid}(b)). This shows that the simple contagion accumulates the critical mass for the complex contagion, which allows other nodes to be infected earlier than when the complex contagion alone takes place. It also implies that the simple and complex contagion may demonstrate identical spreading dynamics under certain parameters.

\section{Conclusion and discussion}
To summarize, we propose a GLT model for the continuous-time complex contagion process. It overcomes the limitations of the LT model in studying the system evolution. The GLT model can be efficiently implemented by the Gillespie Algorithm. We find that the traditional LT model tends to underestimate the speed of spreading and the randomness of the spreading sequence, compared with cases when the dynamics are more properly defined. We show that the GLT model can be very efficient to simulate more complicated spreading. Taken together, the GLT model we proposed can be a useful tool to study complex contagion, especially when the time evolution of the spreading is the focus. Our result not only sheds light on a series of important questions that were not emphasized previously, but also brings insight into the modeling process of real spreading data. Previous research shows that real spreading process is usually more complex. There are examples of combining multiple spreading mechanisms \cite{wang2019anomalous, wu2018model}. More importantly, the recovery process is included in real spreading \cite{hu2018local}. This urges us to combine the linear threshold model with SIR model, which is readily doable with the GLT model proposed in this paper. These more sophisticated models together with real spreading data would definitely help us understand the underlying patterns in information spreading. 

Our model also has some shortcomings, the events based on the spreading dynamics are described as a Poisson random process for the GLT model, which may not deal with the real information spreading well. In the GLT model, whether a node becomes active depends only on the number of current exposures from its neighbors, without memory effects. The previous records, however, could impact the information spreading in current time in the real data \cite{lu2011small}. Miller \cite{miller2016equivalence} studied the equivalence between the generalized epidemic process and the LT model through the percolation theory. They find that the generalized epidemic process is completely equivalent to the LT model. Using the GLT model, we can extend the analyses to the continuous-time dynamics. In the future, we can study the equivalence based on the temporal dynamics between the GLT model and the simple contagion. In addition, we can study under what circumstances the two models can be distinguished, and factors that make the two models equivalent.

\section*{Acknowledgments}
This research is supported by the Chongqing Graduate Research and Innovation Project (Grant No. CYB18080), and the S-Tech Internet Communication Academic Support Plan.

\section*{Data Availability Statement}
Data sharing is not applicable to this article as data were generated by the theoretical model.

\section*{References}
%\bibliography{mybibfile}

\begin{thebibliography}{0}%
\makeatletter
\providecommand \@ifxundefined [1]{%
 \@ifx{#1\undefined}
}%
\providecommand \@ifnum [1]{%
 \ifnum #1\expandafter \@firstoftwo
 \else \expandafter \@secondoftwo
 \fi
}%
\providecommand \@ifx [1]{%
 \ifx #1\expandafter \@firstoftwo
 \else \expandafter \@secondoftwo
 \fi
}%
\providecommand \natexlab [1]{#1}%
\providecommand \enquote  [1]{``#1''}%
\providecommand \bibnamefont  [1]{#1}%
\providecommand \bibfnamefont [1]{#1}%
\providecommand \citenamefont [1]{#1}%
\providecommand \href@noop [0]{\@secondoftwo}%
\providecommand \href [0]{\begingroup \@sanitize@url \@href}%
\providecommand \@href[1]{\@@startlink{#1}\@@href}%
\providecommand \@@href[1]{\endgroup#1\@@endlink}%
\providecommand \@sanitize@url [0]{\catcode `\\12\catcode `\$12\catcode
  `\&12\catcode `\#12\catcode `\^12\catcode `\_12\catcode `\%12\relax}%
\providecommand \@@startlink[1]{}%
\providecommand \@@endlink[0]{}%
\providecommand \url  [0]{\begingroup\@sanitize@url \@url }%
\providecommand \@url [1]{\endgroup\@href {#1}{\urlprefix }}%
\providecommand \urlprefix  [0]{URL }%
\providecommand \Eprint [0]{\href }%
\providecommand \doibase [0]{http://dx.doi.org/}%
\providecommand \selectlanguage [0]{\@gobble}%
\providecommand \bibinfo  [0]{\@secondoftwo}%
\providecommand \bibfield  [0]{\@secondoftwo}%
\providecommand \translation [1]{[#1]}%
\providecommand \BibitemOpen [0]{}%
\providecommand \bibitemStop [0]{}%
\providecommand \bibitemNoStop [0]{.\EOS\space}%
\providecommand \EOS [0]{\spacefactor3000\relax}%
\providecommand \BibitemShut  [1]{\csname bibitem#1\endcsname}%
\let\auto@bib@innerbib\@empty
%</preamble>
\end{thebibliography}%


\begin{thebibliography}{59}%
\makeatletter
\providecommand \@ifxundefined [1]{%
 \@ifx{#1\undefined}
}%
\providecommand \@ifnum [1]{%
 \ifnum #1\expandafter \@firstoftwo
 \else \expandafter \@secondoftwo
 \fi
}%
\providecommand \@ifx [1]{%
 \ifx #1\expandafter \@firstoftwo
 \else \expandafter \@secondoftwo
 \fi
}%
\providecommand \natexlab [1]{#1}%
\providecommand \enquote  [1]{``#1''}%
\providecommand \bibnamefont  [1]{#1}%
\providecommand \bibfnamefont [1]{#1}%
\providecommand \citenamefont [1]{#1}%
\providecommand \href@noop [0]{\@secondoftwo}%
\providecommand \href [0]{\begingroup \@sanitize@url \@href}%
\providecommand \@href[1]{\@@startlink{#1}\@@href}%
\providecommand \@@href[1]{\endgroup#1\@@endlink}%
\providecommand \@sanitize@url [0]{\catcode `\\12\catcode `\$12\catcode
  `\&12\catcode `\#12\catcode `\^12\catcode `\_12\catcode `\%12\relax}%
\providecommand \@@startlink[1]{}%
\providecommand \@@endlink[0]{}%
\providecommand \url  [0]{\begingroup\@sanitize@url \@url }%
\providecommand \@url [1]{\endgroup\@href {#1}{\urlprefix }}%
\providecommand \urlprefix  [0]{URL }%
\providecommand \Eprint [0]{\href }%
\providecommand \doibase [0]{http://dx.doi.org/}%
\providecommand \selectlanguage [0]{\@gobble}%
\providecommand \bibinfo  [0]{\@secondoftwo}%
\providecommand \bibfield  [0]{\@secondoftwo}%
\providecommand \translation [1]{[#1]}%
\providecommand \BibitemOpen [0]{}%
\providecommand \bibitemStop [0]{}%
\providecommand \bibitemNoStop [0]{.\EOS\space}%
\providecommand \EOS [0]{\spacefactor3000\relax}%
\providecommand \BibitemShut  [1]{\csname bibitem#1\endcsname}%
\let\auto@bib@innerbib\@empty
%</preamble>
\bibitem [{\citenamefont {Rogers}(2010)}]{rogers2010diffusion}%
  \BibitemOpen
  \bibfield  {author} {\bibinfo {author} {\bibfnamefont {E.~M.}\ \bibnamefont
  {Rogers}},\ }\href@noop {} {\emph {\bibinfo {title} {Diffusion of
  innovations}}}\ (\bibinfo  {publisher} {Simon and Schuster},\ \bibinfo {year}
  {2010})\BibitemShut {NoStop}%
\bibitem [{\citenamefont {Weiss}\ \emph {et~al.}(2014)\citenamefont {Weiss},
  \citenamefont {Poncela-Casasnovas}, \citenamefont {Glaser}, \citenamefont
  {Pah}, \citenamefont {Persell}, \citenamefont {Baker}, \citenamefont
  {Wunderink},\ and\ \citenamefont {Amaral}}]{weiss2014adoption}%
  \BibitemOpen
  \bibfield  {author} {\bibinfo {author} {\bibfnamefont {C.~H.}\ \bibnamefont
  {Weiss}}, \bibinfo {author} {\bibfnamefont {J.}~\bibnamefont
  {Poncela-Casasnovas}}, \bibinfo {author} {\bibfnamefont {J.~I.}\ \bibnamefont
  {Glaser}}, \bibinfo {author} {\bibfnamefont {A.~R.}\ \bibnamefont {Pah}},
  \bibinfo {author} {\bibfnamefont {S.~D.}\ \bibnamefont {Persell}}, \bibinfo
  {author} {\bibfnamefont {D.~W.}\ \bibnamefont {Baker}}, \bibinfo {author}
  {\bibfnamefont {R.~G.}\ \bibnamefont {Wunderink}}, \ and\ \bibinfo {author}
  {\bibfnamefont {L.~A.~N.}\ \bibnamefont {Amaral}},\ }\bibfield  {title}
  {\enquote {\bibinfo {title} {Adoption of a high-impact innovation in a
  homogeneous population},}\ }\href@noop {} {\bibfield  {journal} {\bibinfo
  {journal} {Physical review x}\ }\textbf {\bibinfo {volume} {4}},\ \bibinfo
  {pages} {041008} (\bibinfo {year} {2014})}\BibitemShut {NoStop}%
\bibitem [{\citenamefont {Zhang}\ \emph {et~al.}(2016)\citenamefont {Zhang},
  \citenamefont {Liu}, \citenamefont {Zhan}, \citenamefont {Lu}, \citenamefont
  {Zhang},\ and\ \citenamefont {Zhang}}]{zhang2016dynamics}%
  \BibitemOpen
  \bibfield  {author} {\bibinfo {author} {\bibfnamefont {Z.-K.}\ \bibnamefont
  {Zhang}}, \bibinfo {author} {\bibfnamefont {C.}~\bibnamefont {Liu}}, \bibinfo
  {author} {\bibfnamefont {X.-X.}\ \bibnamefont {Zhan}}, \bibinfo {author}
  {\bibfnamefont {X.}~\bibnamefont {Lu}}, \bibinfo {author} {\bibfnamefont
  {C.-X.}\ \bibnamefont {Zhang}}, \ and\ \bibinfo {author} {\bibfnamefont
  {Y.-C.}\ \bibnamefont {Zhang}},\ }\bibfield  {title} {\enquote {\bibinfo
  {title} {Dynamics of information diffusion and its applications on complex
  networks},}\ }\href@noop {} {\bibfield  {journal} {\bibinfo  {journal}
  {Physics Reports}\ }\textbf {\bibinfo {volume} {651}},\ \bibinfo {pages}
  {1--34} (\bibinfo {year} {2016})}\BibitemShut {NoStop}%
\bibitem [{\citenamefont {Bass}(1969)}]{bass1969new}%
  \BibitemOpen
  \bibfield  {author} {\bibinfo {author} {\bibfnamefont {F.~M.}\ \bibnamefont
  {Bass}},\ }\bibfield  {title} {\enquote {\bibinfo {title} {A new product
  growth for model consumer durables},}\ }\href@noop {} {\bibfield  {journal}
  {\bibinfo  {journal} {Management science}\ }\textbf {\bibinfo {volume}
  {15}},\ \bibinfo {pages} {215--227} (\bibinfo {year} {1969})}\BibitemShut
  {NoStop}%
\bibitem [{\citenamefont {Aral}, \citenamefont {Muchnik},\ and\ \citenamefont
  {Sundararajan}(2009)}]{aral2009distinguishing}%
  \BibitemOpen
  \bibfield  {author} {\bibinfo {author} {\bibfnamefont {S.}~\bibnamefont
  {Aral}}, \bibinfo {author} {\bibfnamefont {L.}~\bibnamefont {Muchnik}}, \
  and\ \bibinfo {author} {\bibfnamefont {A.}~\bibnamefont {Sundararajan}},\
  }\bibfield  {title} {\enquote {\bibinfo {title} {Distinguishing
  influence-based contagion from homophily-driven diffusion in dynamic
  networks},}\ }\href@noop {} {\bibfield  {journal} {\bibinfo  {journal}
  {Proceedings of the National Academy of Sciences}\ }\textbf {\bibinfo
  {volume} {106}},\ \bibinfo {pages} {21544--21549} (\bibinfo {year}
  {2009})}\BibitemShut {NoStop}%
\bibitem [{\citenamefont {Jin}\ \emph {et~al.}(2019)\citenamefont {Jin},
  \citenamefont {Song}, \citenamefont {Bjelland}, \citenamefont {Canright},\
  and\ \citenamefont {Wang}}]{jin2019emergence}%
  \BibitemOpen
  \bibfield  {author} {\bibinfo {author} {\bibfnamefont {C.}~\bibnamefont
  {Jin}}, \bibinfo {author} {\bibfnamefont {C.}~\bibnamefont {Song}}, \bibinfo
  {author} {\bibfnamefont {J.}~\bibnamefont {Bjelland}}, \bibinfo {author}
  {\bibfnamefont {G.}~\bibnamefont {Canright}}, \ and\ \bibinfo {author}
  {\bibfnamefont {D.}~\bibnamefont {Wang}},\ }\bibfield  {title} {\enquote
  {\bibinfo {title} {Emergence of scaling in complex substitutive systems},}\
  }\href@noop {} {\bibfield  {journal} {\bibinfo  {journal} {Nature human
  behaviour}\ }\textbf {\bibinfo {volume} {3}},\ \bibinfo {pages} {837--846}
  (\bibinfo {year} {2019})}\BibitemShut {NoStop}%
\bibitem [{\citenamefont {Fowler}\ and\ \citenamefont
  {Christakis}(2010)}]{fowler2010cooperative}%
  \BibitemOpen
  \bibfield  {author} {\bibinfo {author} {\bibfnamefont {J.~H.}\ \bibnamefont
  {Fowler}}\ and\ \bibinfo {author} {\bibfnamefont {N.~A.}\ \bibnamefont
  {Christakis}},\ }\bibfield  {title} {\enquote {\bibinfo {title} {Cooperative
  behavior cascades in human social networks},}\ }\href@noop {} {\bibfield
  {journal} {\bibinfo  {journal} {Proceedings of the National Academy of
  Sciences}\ }\textbf {\bibinfo {volume} {107}},\ \bibinfo {pages} {5334--5338}
  (\bibinfo {year} {2010})}\BibitemShut {NoStop}%
\bibitem [{\citenamefont {Zheng}\ \emph {et~al.}(2013)\citenamefont {Zheng},
  \citenamefont {L{\"u}}, \citenamefont {Zhao} \emph
  {et~al.}}]{zheng2013spreading}%
  \BibitemOpen
  \bibfield  {author} {\bibinfo {author} {\bibfnamefont {M.}~\bibnamefont
  {Zheng}}, \bibinfo {author} {\bibfnamefont {L.}~\bibnamefont {L{\"u}}},
  \bibinfo {author} {\bibfnamefont {M.}~\bibnamefont {Zhao}},  \emph {et~al.},\
  }\bibfield  {title} {\enquote {\bibinfo {title} {Spreading in online social
  networks: The role of social reinforcement},}\ }\href@noop {} {\bibfield
  {journal} {\bibinfo  {journal} {Physical Review E}\ }\textbf {\bibinfo
  {volume} {88}},\ \bibinfo {pages} {012818} (\bibinfo {year}
  {2013})}\BibitemShut {NoStop}%
\bibitem [{\citenamefont {Jia}, \citenamefont {Wang},\ and\ \citenamefont
  {Szymanski}(2017)}]{jia2017quantifying}%
  \BibitemOpen
  \bibfield  {author} {\bibinfo {author} {\bibfnamefont {T.}~\bibnamefont
  {Jia}}, \bibinfo {author} {\bibfnamefont {D.}~\bibnamefont {Wang}}, \ and\
  \bibinfo {author} {\bibfnamefont {B.~K.}\ \bibnamefont {Szymanski}},\
  }\bibfield  {title} {\enquote {\bibinfo {title} {Quantifying patterns of
  research-interest evolution},}\ }\href@noop {} {\bibfield  {journal}
  {\bibinfo  {journal} {Nature Human Behaviour}\ }\textbf {\bibinfo {volume}
  {1}},\ \bibinfo {pages} {0078} (\bibinfo {year} {2017})}\BibitemShut
  {NoStop}%
\bibitem [{\citenamefont {Moreno}, \citenamefont {Nekovee},\ and\ \citenamefont
  {Pacheco}(2004)}]{moreno2004dynamics}%
  \BibitemOpen
  \bibfield  {author} {\bibinfo {author} {\bibfnamefont {Y.}~\bibnamefont
  {Moreno}}, \bibinfo {author} {\bibfnamefont {M.}~\bibnamefont {Nekovee}}, \
  and\ \bibinfo {author} {\bibfnamefont {A.~F.}\ \bibnamefont {Pacheco}},\
  }\bibfield  {title} {\enquote {\bibinfo {title} {Dynamics of rumor spreading
  in complex networks},}\ }\href@noop {} {\bibfield  {journal} {\bibinfo
  {journal} {Physical Review E}\ }\textbf {\bibinfo {volume} {69}},\ \bibinfo
  {pages} {066130} (\bibinfo {year} {2004})}\BibitemShut {NoStop}%
\bibitem [{\citenamefont {Lazer}\ \emph {et~al.}(2018)\citenamefont {Lazer},
  \citenamefont {Baum}, \citenamefont {Benkler}, \citenamefont {Berinsky},
  \citenamefont {Greenhill}, \citenamefont {Menczer}, \citenamefont {Metzger},
  \citenamefont {Nyhan}, \citenamefont {Pennycook}, \citenamefont {Rothschild}
  \emph {et~al.}}]{lazer2018science}%
  \BibitemOpen
  \bibfield  {author} {\bibinfo {author} {\bibfnamefont {D.~M.}\ \bibnamefont
  {Lazer}}, \bibinfo {author} {\bibfnamefont {M.~A.}\ \bibnamefont {Baum}},
  \bibinfo {author} {\bibfnamefont {Y.}~\bibnamefont {Benkler}}, \bibinfo
  {author} {\bibfnamefont {A.~J.}\ \bibnamefont {Berinsky}}, \bibinfo {author}
  {\bibfnamefont {K.~M.}\ \bibnamefont {Greenhill}}, \bibinfo {author}
  {\bibfnamefont {F.}~\bibnamefont {Menczer}}, \bibinfo {author} {\bibfnamefont
  {M.~J.}\ \bibnamefont {Metzger}}, \bibinfo {author} {\bibfnamefont
  {B.}~\bibnamefont {Nyhan}}, \bibinfo {author} {\bibfnamefont
  {G.}~\bibnamefont {Pennycook}}, \bibinfo {author} {\bibfnamefont
  {D.}~\bibnamefont {Rothschild}},  \emph {et~al.},\ }\bibfield  {title}
  {\enquote {\bibinfo {title} {The science of fake news},}\ }\href@noop {}
  {\bibfield  {journal} {\bibinfo  {journal} {Science}\ }\textbf {\bibinfo
  {volume} {359}},\ \bibinfo {pages} {1094--1096} (\bibinfo {year}
  {2018})}\BibitemShut {NoStop}%
\bibitem [{\citenamefont {Vosoughi}, \citenamefont {Roy},\ and\ \citenamefont
  {Aral}(2018)}]{vosoughi2018spread}%
  \BibitemOpen
  \bibfield  {author} {\bibinfo {author} {\bibfnamefont {S.}~\bibnamefont
  {Vosoughi}}, \bibinfo {author} {\bibfnamefont {D.}~\bibnamefont {Roy}}, \
  and\ \bibinfo {author} {\bibfnamefont {S.}~\bibnamefont {Aral}},\ }\bibfield
  {title} {\enquote {\bibinfo {title} {The spread of true and false news
  online},}\ }\href@noop {} {\bibfield  {journal} {\bibinfo  {journal}
  {Science}\ }\textbf {\bibinfo {volume} {359}},\ \bibinfo {pages} {1146--1151}
  (\bibinfo {year} {2018})}\BibitemShut {NoStop}%
\bibitem [{\citenamefont {Travieso}\ and\ \citenamefont
  {da~Fontoura~Costa}(2006)}]{travieso2006spread}%
  \BibitemOpen
  \bibfield  {author} {\bibinfo {author} {\bibfnamefont {G.}~\bibnamefont
  {Travieso}}\ and\ \bibinfo {author} {\bibfnamefont {L.}~\bibnamefont
  {da~Fontoura~Costa}},\ }\bibfield  {title} {\enquote {\bibinfo {title}
  {Spread of opinions and proportional voting},}\ }\href@noop {} {\bibfield
  {journal} {\bibinfo  {journal} {Physical Review E}\ }\textbf {\bibinfo
  {volume} {74}},\ \bibinfo {pages} {036112} (\bibinfo {year}
  {2006})}\BibitemShut {NoStop}%
\bibitem [{\citenamefont {Battiston}\ \emph {et~al.}(2020)\citenamefont
  {Battiston}, \citenamefont {Cencetti}, \citenamefont {Iacopini},
  \citenamefont {Latora}, \citenamefont {Lucas}, \citenamefont {Patania},
  \citenamefont {Young},\ and\ \citenamefont {Petri}}]{battiston2020networks}%
  \BibitemOpen
  \bibfield  {author} {\bibinfo {author} {\bibfnamefont {F.}~\bibnamefont
  {Battiston}}, \bibinfo {author} {\bibfnamefont {G.}~\bibnamefont {Cencetti}},
  \bibinfo {author} {\bibfnamefont {I.}~\bibnamefont {Iacopini}}, \bibinfo
  {author} {\bibfnamefont {V.}~\bibnamefont {Latora}}, \bibinfo {author}
  {\bibfnamefont {M.}~\bibnamefont {Lucas}}, \bibinfo {author} {\bibfnamefont
  {A.}~\bibnamefont {Patania}}, \bibinfo {author} {\bibfnamefont {J.-G.}\
  \bibnamefont {Young}}, \ and\ \bibinfo {author} {\bibfnamefont
  {G.}~\bibnamefont {Petri}},\ }\bibfield  {title} {\enquote {\bibinfo {title}
  {Networks beyond pairwise interactions: structure and dynamics},}\
  }\href@noop {} {\bibfield  {journal} {\bibinfo  {journal} {Physics Reports}\
  } (\bibinfo {year} {2020})}\BibitemShut {NoStop}%
\bibitem [{\citenamefont {Evans}\ and\ \citenamefont
  {Foster}(2011)}]{evans2011metaknowledge}%
  \BibitemOpen
  \bibfield  {author} {\bibinfo {author} {\bibfnamefont {J.~A.}\ \bibnamefont
  {Evans}}\ and\ \bibinfo {author} {\bibfnamefont {J.~G.}\ \bibnamefont
  {Foster}},\ }\bibfield  {title} {\enquote {\bibinfo {title}
  {Metaknowledge},}\ }\href@noop {} {\bibfield  {journal} {\bibinfo  {journal}
  {Science}\ }\textbf {\bibinfo {volume} {331}},\ \bibinfo {pages} {721--725}
  (\bibinfo {year} {2011})}\BibitemShut {NoStop}%
\bibitem [{\citenamefont {Iacopini}, \citenamefont {Milojevi{\'c}},\ and\
  \citenamefont {Latora}(2018)}]{iacopini2018network}%
  \BibitemOpen
  \bibfield  {author} {\bibinfo {author} {\bibfnamefont {I.}~\bibnamefont
  {Iacopini}}, \bibinfo {author} {\bibfnamefont {S.}~\bibnamefont
  {Milojevi{\'c}}}, \ and\ \bibinfo {author} {\bibfnamefont {V.}~\bibnamefont
  {Latora}},\ }\bibfield  {title} {\enquote {\bibinfo {title} {Network dynamics
  of innovation processes},}\ }\href@noop {} {\bibfield  {journal} {\bibinfo
  {journal} {Physical review letters}\ }\textbf {\bibinfo {volume} {120}},\
  \bibinfo {pages} {048301} (\bibinfo {year} {2018})}\BibitemShut {NoStop}%
\bibitem [{\citenamefont {Liu}\ \emph {et~al.}(2014)\citenamefont {Liu},
  \citenamefont {Perra}, \citenamefont {Karsai},\ and\ \citenamefont
  {Vespignani}}]{liu2014controlling}%
  \BibitemOpen
  \bibfield  {author} {\bibinfo {author} {\bibfnamefont {S.}~\bibnamefont
  {Liu}}, \bibinfo {author} {\bibfnamefont {N.}~\bibnamefont {Perra}}, \bibinfo
  {author} {\bibfnamefont {M.}~\bibnamefont {Karsai}}, \ and\ \bibinfo {author}
  {\bibfnamefont {A.}~\bibnamefont {Vespignani}},\ }\bibfield  {title}
  {\enquote {\bibinfo {title} {Controlling contagion processes in activity
  driven networks},}\ }\href@noop {} {\bibfield  {journal} {\bibinfo  {journal}
  {Physical review letters}\ }\textbf {\bibinfo {volume} {112}},\ \bibinfo
  {pages} {118702} (\bibinfo {year} {2014})}\BibitemShut {NoStop}%
\bibitem [{\citenamefont {Guilbeault}, \citenamefont {Becker},\ and\
  \citenamefont {Centola}(2018)}]{guilbeault2018complex}%
  \BibitemOpen
  \bibfield  {author} {\bibinfo {author} {\bibfnamefont {D.}~\bibnamefont
  {Guilbeault}}, \bibinfo {author} {\bibfnamefont {J.}~\bibnamefont {Becker}},
  \ and\ \bibinfo {author} {\bibfnamefont {D.}~\bibnamefont {Centola}},\
  }\bibfield  {title} {\enquote {\bibinfo {title} {Complex contagions: A decade
  in review},}\ }in\ \href@noop {} {\emph {\bibinfo {booktitle} {Complex
  spreading phenomena in social systems}}}\ (\bibinfo  {publisher} {Springer},\
  \bibinfo {year} {2018})\ pp.\ \bibinfo {pages} {3--25}\BibitemShut {NoStop}%
\bibitem [{\citenamefont {Centola}(2018)}]{centola2018behavior}%
  \BibitemOpen
  \bibfield  {author} {\bibinfo {author} {\bibfnamefont {D.}~\bibnamefont
  {Centola}},\ }\href@noop {} {\emph {\bibinfo {title} {How behavior spreads:
  The science of complex contagions}}},\ Vol.~\bibinfo {volume} {3}\ (\bibinfo
  {publisher} {Princeton University Press},\ \bibinfo {year}
  {2018})\BibitemShut {NoStop}%
\bibitem [{\citenamefont {Centola}(2010)}]{centola2010spread}%
  \BibitemOpen
  \bibfield  {author} {\bibinfo {author} {\bibfnamefont {D.}~\bibnamefont
  {Centola}},\ }\bibfield  {title} {\enquote {\bibinfo {title} {The spread of
  behavior in an online social network experiment},}\ }\href@noop {} {\bibfield
   {journal} {\bibinfo  {journal} {science}\ }\textbf {\bibinfo {volume}
  {329}},\ \bibinfo {pages} {1194--1197} (\bibinfo {year} {2010})}\BibitemShut
  {NoStop}%
\bibitem [{\citenamefont {Wang}\ \emph {et~al.}(2019)\citenamefont {Wang},
  \citenamefont {Liu}, \citenamefont {Liang}, \citenamefont {Hu},\ and\
  \citenamefont {Zhou}}]{wang2019coevolution}%
  \BibitemOpen
  \bibfield  {author} {\bibinfo {author} {\bibfnamefont {W.}~\bibnamefont
  {Wang}}, \bibinfo {author} {\bibfnamefont {Q.-H.}\ \bibnamefont {Liu}},
  \bibinfo {author} {\bibfnamefont {J.}~\bibnamefont {Liang}}, \bibinfo
  {author} {\bibfnamefont {Y.}~\bibnamefont {Hu}}, \ and\ \bibinfo {author}
  {\bibfnamefont {T.}~\bibnamefont {Zhou}},\ }\bibfield  {title} {\enquote
  {\bibinfo {title} {Coevolution spreading in complex networks},}\ }\href@noop
  {} {\bibfield  {journal} {\bibinfo  {journal} {Physics Reports}\ } (\bibinfo
  {year} {2019})}\BibitemShut {NoStop}%
\bibitem [{\citenamefont {Karsai}\ \emph {et~al.}(2011)\citenamefont {Karsai},
  \citenamefont {Kivel{\"a}}, \citenamefont {Pan}, \citenamefont {Kaski},
  \citenamefont {Kert{\'e}sz}, \citenamefont {Barab{\'a}si},\ and\
  \citenamefont {Saram{\"a}ki}}]{karsai2011small}%
  \BibitemOpen
  \bibfield  {author} {\bibinfo {author} {\bibfnamefont {M.}~\bibnamefont
  {Karsai}}, \bibinfo {author} {\bibfnamefont {M.}~\bibnamefont {Kivel{\"a}}},
  \bibinfo {author} {\bibfnamefont {R.~K.}\ \bibnamefont {Pan}}, \bibinfo
  {author} {\bibfnamefont {K.}~\bibnamefont {Kaski}}, \bibinfo {author}
  {\bibfnamefont {J.}~\bibnamefont {Kert{\'e}sz}}, \bibinfo {author}
  {\bibfnamefont {A.-L.}\ \bibnamefont {Barab{\'a}si}}, \ and\ \bibinfo
  {author} {\bibfnamefont {J.}~\bibnamefont {Saram{\"a}ki}},\ }\bibfield
  {title} {\enquote {\bibinfo {title} {Small but slow world: How network
  topology and burstiness slow down spreading},}\ }\href@noop {} {\bibfield
  {journal} {\bibinfo  {journal} {Physical Review E}\ }\textbf {\bibinfo
  {volume} {83}},\ \bibinfo {pages} {025102} (\bibinfo {year}
  {2011})}\BibitemShut {NoStop}%
\bibitem [{\citenamefont {L{\"u}}, \citenamefont {Chen},\ and\ \citenamefont
  {Zhou}(2011)}]{lu2011small}%
  \BibitemOpen
  \bibfield  {author} {\bibinfo {author} {\bibfnamefont {L.}~\bibnamefont
  {L{\"u}}}, \bibinfo {author} {\bibfnamefont {D.-B.}\ \bibnamefont {Chen}}, \
  and\ \bibinfo {author} {\bibfnamefont {T.}~\bibnamefont {Zhou}},\ }\bibfield
  {title} {\enquote {\bibinfo {title} {The small world yields the most
  effective information spreading},}\ }\href@noop {} {\bibfield  {journal}
  {\bibinfo  {journal} {New Journal of Physics}\ }\textbf {\bibinfo {volume}
  {13}},\ \bibinfo {pages} {123005} (\bibinfo {year} {2011})}\BibitemShut
  {NoStop}%
\bibitem [{\citenamefont {Xian}\ \emph {et~al.}(2019)\citenamefont {Xian},
  \citenamefont {Yang}, \citenamefont {Pan}, \citenamefont {Wang},\ and\
  \citenamefont {Wang}}]{xian2019misinformation}%
  \BibitemOpen
  \bibfield  {author} {\bibinfo {author} {\bibfnamefont {J.}~\bibnamefont
  {Xian}}, \bibinfo {author} {\bibfnamefont {D.}~\bibnamefont {Yang}}, \bibinfo
  {author} {\bibfnamefont {L.}~\bibnamefont {Pan}}, \bibinfo {author}
  {\bibfnamefont {W.}~\bibnamefont {Wang}}, \ and\ \bibinfo {author}
  {\bibfnamefont {Z.}~\bibnamefont {Wang}},\ }\bibfield  {title} {\enquote
  {\bibinfo {title} {Misinformation spreading on correlated multiplex
  networks},}\ }\href@noop {} {\bibfield  {journal} {\bibinfo  {journal}
  {Chaos: An Interdisciplinary Journal of Nonlinear Science}\ }\textbf
  {\bibinfo {volume} {29}},\ \bibinfo {pages} {113123} (\bibinfo {year}
  {2019})}\BibitemShut {NoStop}%
\bibitem [{\citenamefont {Castellano}\ and\ \citenamefont
  {Pastor-Satorras}(2010)}]{castellano2010thresholds}%
  \BibitemOpen
  \bibfield  {author} {\bibinfo {author} {\bibfnamefont {C.}~\bibnamefont
  {Castellano}}\ and\ \bibinfo {author} {\bibfnamefont {R.}~\bibnamefont
  {Pastor-Satorras}},\ }\bibfield  {title} {\enquote {\bibinfo {title}
  {Thresholds for epidemic spreading in networks},}\ }\href@noop {} {\bibfield
  {journal} {\bibinfo  {journal} {Physical review letters}\ }\textbf {\bibinfo
  {volume} {105}},\ \bibinfo {pages} {218701} (\bibinfo {year}
  {2010})}\BibitemShut {NoStop}%
\bibitem [{\citenamefont {Borge-Holthoefer}\ \emph {et~al.}(2013)\citenamefont
  {Borge-Holthoefer}, \citenamefont {Ba{\~n}os}, \citenamefont
  {Gonz{\'a}lez-Bail{\'o}n},\ and\ \citenamefont
  {Moreno}}]{borge2013cascading}%
  \BibitemOpen
  \bibfield  {author} {\bibinfo {author} {\bibfnamefont {J.}~\bibnamefont
  {Borge-Holthoefer}}, \bibinfo {author} {\bibfnamefont {R.~A.}\ \bibnamefont
  {Ba{\~n}os}}, \bibinfo {author} {\bibfnamefont {S.}~\bibnamefont
  {Gonz{\'a}lez-Bail{\'o}n}}, \ and\ \bibinfo {author} {\bibfnamefont
  {Y.}~\bibnamefont {Moreno}},\ }\bibfield  {title} {\enquote {\bibinfo {title}
  {Cascading behaviour in complex socio-technical networks},}\ }\href@noop {}
  {\bibfield  {journal} {\bibinfo  {journal} {Journal of Complex Networks}\
  }\textbf {\bibinfo {volume} {1}},\ \bibinfo {pages} {3--24} (\bibinfo {year}
  {2013})}\BibitemShut {NoStop}%
\bibitem [{\citenamefont {Watts}(2002)}]{watts2002simple}%
  \BibitemOpen
  \bibfield  {author} {\bibinfo {author} {\bibfnamefont {D.~J.}\ \bibnamefont
  {Watts}},\ }\bibfield  {title} {\enquote {\bibinfo {title} {A simple model of
  global cascades on random networks},}\ }\href@noop {} {\bibfield  {journal}
  {\bibinfo  {journal} {Proceedings of the National Academy of Sciences}\
  }\textbf {\bibinfo {volume} {99}},\ \bibinfo {pages} {5766--5771} (\bibinfo
  {year} {2002})}\BibitemShut {NoStop}%
\bibitem [{\citenamefont {Karimi}\ and\ \citenamefont
  {Holme}(2013)}]{karimi2013threshold}%
  \BibitemOpen
  \bibfield  {author} {\bibinfo {author} {\bibfnamefont {F.}~\bibnamefont
  {Karimi}}\ and\ \bibinfo {author} {\bibfnamefont {P.}~\bibnamefont {Holme}},\
  }\bibfield  {title} {\enquote {\bibinfo {title} {Threshold model of cascades
  in empirical temporal networks},}\ }\href@noop {} {\bibfield  {journal}
  {\bibinfo  {journal} {Physica A: Statistical Mechanics and its Applications}\
  }\textbf {\bibinfo {volume} {392}},\ \bibinfo {pages} {3476--3483} (\bibinfo
  {year} {2013})}\BibitemShut {NoStop}%
\bibitem [{\citenamefont {Wang}\ \emph {et~al.}(2016)\citenamefont {Wang},
  \citenamefont {Tang}, \citenamefont {Shu},\ and\ \citenamefont
  {Wang}}]{wang2016dynamics}%
  \BibitemOpen
  \bibfield  {author} {\bibinfo {author} {\bibfnamefont {W.}~\bibnamefont
  {Wang}}, \bibinfo {author} {\bibfnamefont {M.}~\bibnamefont {Tang}}, \bibinfo
  {author} {\bibfnamefont {P.}~\bibnamefont {Shu}}, \ and\ \bibinfo {author}
  {\bibfnamefont {Z.}~\bibnamefont {Wang}},\ }\bibfield  {title} {\enquote
  {\bibinfo {title} {Dynamics of social contagions with heterogeneous adoption
  thresholds: crossover phenomena in phase transition},}\ }\href@noop {}
  {\bibfield  {journal} {\bibinfo  {journal} {New Journal of Physics}\ }\textbf
  {\bibinfo {volume} {18}},\ \bibinfo {pages} {013029} (\bibinfo {year}
  {2016})}\BibitemShut {NoStop}%
\bibitem [{\citenamefont {Granovetter}(1978)}]{granovetter1978threshold}%
  \BibitemOpen
  \bibfield  {author} {\bibinfo {author} {\bibfnamefont {M.}~\bibnamefont
  {Granovetter}},\ }\bibfield  {title} {\enquote {\bibinfo {title} {Threshold
  models of collective behavior},}\ }\href@noop {} {\bibfield  {journal}
  {\bibinfo  {journal} {American journal of sociology}\ }\textbf {\bibinfo
  {volume} {83}},\ \bibinfo {pages} {1420--1443} (\bibinfo {year}
  {1978})}\BibitemShut {NoStop}%
\bibitem [{\citenamefont {Kempe}, \citenamefont {Kleinberg},\ and\
  \citenamefont {Tardos}(2003)}]{kempe2003maximizing}%
  \BibitemOpen
  \bibfield  {author} {\bibinfo {author} {\bibfnamefont {D.}~\bibnamefont
  {Kempe}}, \bibinfo {author} {\bibfnamefont {J.}~\bibnamefont {Kleinberg}}, \
  and\ \bibinfo {author} {\bibfnamefont {{\'E}.}~\bibnamefont {Tardos}},\
  }\bibfield  {title} {\enquote {\bibinfo {title} {Maximizing the spread of
  influence through a social network},}\ }in\ \href@noop {} {\emph {\bibinfo
  {booktitle} {Proceedings of the ninth ACM SIGKDD international conference on
  Knowledge discovery and data mining}}}\ (\bibinfo {organization} {ACM},\
  \bibinfo {year} {2003})\ pp.\ \bibinfo {pages} {137--146}\BibitemShut
  {NoStop}%
\bibitem [{\citenamefont {Singh}\ \emph {et~al.}(2013)\citenamefont {Singh},
  \citenamefont {Sreenivasan}, \citenamefont {Szymanski},\ and\ \citenamefont
  {Korniss}}]{singh2013threshold}%
  \BibitemOpen
  \bibfield  {author} {\bibinfo {author} {\bibfnamefont {P.}~\bibnamefont
  {Singh}}, \bibinfo {author} {\bibfnamefont {S.}~\bibnamefont {Sreenivasan}},
  \bibinfo {author} {\bibfnamefont {B.~K.}\ \bibnamefont {Szymanski}}, \ and\
  \bibinfo {author} {\bibfnamefont {G.}~\bibnamefont {Korniss}},\ }\bibfield
  {title} {\enquote {\bibinfo {title} {Threshold-limited spreading in social
  networks with multiple initiators},}\ }\href@noop {} {\bibfield  {journal}
  {\bibinfo  {journal} {Scientific reports}\ }\textbf {\bibinfo {volume} {3}},\
  \bibinfo {pages} {2330} (\bibinfo {year} {2013})}\BibitemShut {NoStop}%
\bibitem [{\citenamefont {Liu}\ \emph {et~al.}(2018{\natexlab{a}})\citenamefont
  {Liu}, \citenamefont {L{\"u}}, \citenamefont {Zhang}, \citenamefont {Tang},\
  and\ \citenamefont {Zhou}}]{liu2018impacts}%
  \BibitemOpen
  \bibfield  {author} {\bibinfo {author} {\bibfnamefont {Q.-H.}\ \bibnamefont
  {Liu}}, \bibinfo {author} {\bibfnamefont {F.-M.}\ \bibnamefont {L{\"u}}},
  \bibinfo {author} {\bibfnamefont {Q.}~\bibnamefont {Zhang}}, \bibinfo
  {author} {\bibfnamefont {M.}~\bibnamefont {Tang}}, \ and\ \bibinfo {author}
  {\bibfnamefont {T.}~\bibnamefont {Zhou}},\ }\bibfield  {title} {\enquote
  {\bibinfo {title} {Impacts of opinion leaders on social contagions},}\
  }\href@noop {} {\bibfield  {journal} {\bibinfo  {journal} {Chaos: An
  Interdisciplinary Journal of Nonlinear Science}\ }\textbf {\bibinfo {volume}
  {28}},\ \bibinfo {pages} {053103} (\bibinfo {year}
  {2018}{\natexlab{a}})}\BibitemShut {NoStop}%
\bibitem [{\citenamefont {Chen}\ \emph {et~al.}(2019)\citenamefont {Chen},
  \citenamefont {Sun}, \citenamefont {Tang}, \citenamefont {Tian},\ and\
  \citenamefont {Xie}}]{chen2019identifying}%
  \BibitemOpen
  \bibfield  {author} {\bibinfo {author} {\bibfnamefont {D.-B.}\ \bibnamefont
  {Chen}}, \bibinfo {author} {\bibfnamefont {H.-L.}\ \bibnamefont {Sun}},
  \bibinfo {author} {\bibfnamefont {Q.}~\bibnamefont {Tang}}, \bibinfo {author}
  {\bibfnamefont {S.-Z.}\ \bibnamefont {Tian}}, \ and\ \bibinfo {author}
  {\bibfnamefont {M.}~\bibnamefont {Xie}},\ }\bibfield  {title} {\enquote
  {\bibinfo {title} {Identifying influential spreaders in complex networks by
  propagation probability dynamics},}\ }\href@noop {} {\bibfield  {journal}
  {\bibinfo  {journal} {Chaos: An Interdisciplinary Journal of Nonlinear
  Science}\ }\textbf {\bibinfo {volume} {29}},\ \bibinfo {pages} {033120}
  (\bibinfo {year} {2019})}\BibitemShut {NoStop}%
\bibitem [{\citenamefont {Cao}\ \emph {et~al.}(2017)\citenamefont {Cao},
  \citenamefont {Shen}, \citenamefont {Cen}, \citenamefont {Ouyang},\ and\
  \citenamefont {Cheng}}]{cao2017deephawkes}%
  \BibitemOpen
  \bibfield  {author} {\bibinfo {author} {\bibfnamefont {Q.}~\bibnamefont
  {Cao}}, \bibinfo {author} {\bibfnamefont {H.}~\bibnamefont {Shen}}, \bibinfo
  {author} {\bibfnamefont {K.}~\bibnamefont {Cen}}, \bibinfo {author}
  {\bibfnamefont {W.}~\bibnamefont {Ouyang}}, \ and\ \bibinfo {author}
  {\bibfnamefont {X.}~\bibnamefont {Cheng}},\ }\bibfield  {title} {\enquote
  {\bibinfo {title} {Deephawkes: Bridging the gap between prediction and
  understanding of information cascades},}\ }in\ \href@noop {} {\emph {\bibinfo
  {booktitle} {Proceedings of the 2017 ACM on Conference on Information and
  Knowledge Management}}}\ (\bibinfo {year} {2017})\ pp.\ \bibinfo {pages}
  {1149--1158}\BibitemShut {NoStop}%
\bibitem [{\citenamefont {Gillespie}(1976)}]{gillespie1976general}%
  \BibitemOpen
  \bibfield  {author} {\bibinfo {author} {\bibfnamefont {D.~T.}\ \bibnamefont
  {Gillespie}},\ }\bibfield  {title} {\enquote {\bibinfo {title} {A general
  method for numerically simulating the stochastic time evolution of coupled
  chemical reactions},}\ }\href@noop {} {\bibfield  {journal} {\bibinfo
  {journal} {Journal of computational physics}\ }\textbf {\bibinfo {volume}
  {22}},\ \bibinfo {pages} {403--434} (\bibinfo {year} {1976})}\BibitemShut
  {NoStop}%
\bibitem [{\citenamefont {Gillespie}(1977)}]{gillespie1977exact}%
  \BibitemOpen
  \bibfield  {author} {\bibinfo {author} {\bibfnamefont {D.~T.}\ \bibnamefont
  {Gillespie}},\ }\bibfield  {title} {\enquote {\bibinfo {title} {Exact
  stochastic simulation of coupled chemical reactions},}\ }\href@noop {}
  {\bibfield  {journal} {\bibinfo  {journal} {The journal of physical
  chemistry}\ }\textbf {\bibinfo {volume} {81}},\ \bibinfo {pages} {2340--2361}
  (\bibinfo {year} {1977})}\BibitemShut {NoStop}%
\bibitem [{\citenamefont {Gleeson}(2011)}]{gleeson2011high}%
  \BibitemOpen
  \bibfield  {author} {\bibinfo {author} {\bibfnamefont {J.~P.}\ \bibnamefont
  {Gleeson}},\ }\bibfield  {title} {\enquote {\bibinfo {title} {High-accuracy
  approximation of binary-state dynamics on networks},}\ }\href@noop {}
  {\bibfield  {journal} {\bibinfo  {journal} {Physical Review Letters}\
  }\textbf {\bibinfo {volume} {107}},\ \bibinfo {pages} {068701} (\bibinfo
  {year} {2011})}\BibitemShut {NoStop}%
\bibitem [{\citenamefont {Gleeson}(2013)}]{gleeson2013binary}%
  \BibitemOpen
  \bibfield  {author} {\bibinfo {author} {\bibfnamefont {J.~P.}\ \bibnamefont
  {Gleeson}},\ }\bibfield  {title} {\enquote {\bibinfo {title} {Binary-state
  dynamics on complex networks: Pair approximation and beyond},}\ }\href@noop
  {} {\bibfield  {journal} {\bibinfo  {journal} {Physical Review X}\ }\textbf
  {\bibinfo {volume} {3}},\ \bibinfo {pages} {021004} (\bibinfo {year}
  {2013})}\BibitemShut {NoStop}%
\bibitem [{\citenamefont {Shen}\ \emph {et~al.}(2016)\citenamefont {Shen},
  \citenamefont {Cao}, \citenamefont {Wang}, \citenamefont {Di},\ and\
  \citenamefont {Stanley}}]{shen2016locating}%
  \BibitemOpen
  \bibfield  {author} {\bibinfo {author} {\bibfnamefont {Z.}~\bibnamefont
  {Shen}}, \bibinfo {author} {\bibfnamefont {S.}~\bibnamefont {Cao}}, \bibinfo
  {author} {\bibfnamefont {W.-X.}\ \bibnamefont {Wang}}, \bibinfo {author}
  {\bibfnamefont {Z.}~\bibnamefont {Di}}, \ and\ \bibinfo {author}
  {\bibfnamefont {H.~E.}\ \bibnamefont {Stanley}},\ }\bibfield  {title}
  {\enquote {\bibinfo {title} {Locating the source of diffusion in complex
  networks by time-reversal backward spreading},}\ }\href@noop {} {\bibfield
  {journal} {\bibinfo  {journal} {Physical Review E}\ }\textbf {\bibinfo
  {volume} {93}},\ \bibinfo {pages} {032301} (\bibinfo {year}
  {2016})}\BibitemShut {NoStop}%
\bibitem [{\citenamefont {Bourigault}, \citenamefont {Lamprier},\ and\
  \citenamefont {Gallinari}(2016)}]{bourigault2016representation}%
  \BibitemOpen
  \bibfield  {author} {\bibinfo {author} {\bibfnamefont {S.}~\bibnamefont
  {Bourigault}}, \bibinfo {author} {\bibfnamefont {S.}~\bibnamefont
  {Lamprier}}, \ and\ \bibinfo {author} {\bibfnamefont {P.}~\bibnamefont
  {Gallinari}},\ }\bibfield  {title} {\enquote {\bibinfo {title}
  {Representation learning for information diffusion through social networks:
  an embedded cascade model},}\ }in\ \href@noop {} {\emph {\bibinfo {booktitle}
  {Proceedings of the Ninth ACM international conference on Web Search and Data
  Mining}}}\ (\bibinfo {year} {2016})\ pp.\ \bibinfo {pages}
  {573--582}\BibitemShut {NoStop}%
\bibitem [{\citenamefont {Gou}\ \emph {et~al.}(2018)\citenamefont {Gou},
  \citenamefont {Shen}, \citenamefont {Du}, \citenamefont {Wu}, \citenamefont
  {Liu},\ and\ \citenamefont {Cheng}}]{gou2018learning}%
  \BibitemOpen
  \bibfield  {author} {\bibinfo {author} {\bibfnamefont {C.}~\bibnamefont
  {Gou}}, \bibinfo {author} {\bibfnamefont {H.}~\bibnamefont {Shen}}, \bibinfo
  {author} {\bibfnamefont {P.}~\bibnamefont {Du}}, \bibinfo {author}
  {\bibfnamefont {D.}~\bibnamefont {Wu}}, \bibinfo {author} {\bibfnamefont
  {Y.}~\bibnamefont {Liu}}, \ and\ \bibinfo {author} {\bibfnamefont
  {X.}~\bibnamefont {Cheng}},\ }\bibfield  {title} {\enquote {\bibinfo {title}
  {Learning sequential features for cascade outbreak prediction},}\ }\href@noop
  {} {\bibfield  {journal} {\bibinfo  {journal} {Knowledge and Information
  Systems}\ }\textbf {\bibinfo {volume} {57}},\ \bibinfo {pages} {721--739}
  (\bibinfo {year} {2018})}\BibitemShut {NoStop}%
\bibitem [{\citenamefont {Porter}\ and\ \citenamefont
  {Gleeson}(2016)}]{porter2016dynamical}%
  \BibitemOpen
  \bibfield  {author} {\bibinfo {author} {\bibfnamefont {M.~A.}\ \bibnamefont
  {Porter}}\ and\ \bibinfo {author} {\bibfnamefont {J.~P.}\ \bibnamefont
  {Gleeson}},\ }\bibfield  {title} {\enquote {\bibinfo {title} {Dynamical
  systems on networks},}\ }\href@noop {} {\bibfield  {journal} {\bibinfo
  {journal} {Frontiers in Applied Dynamical Systems: Reviews and Tutorials}\
  }\textbf {\bibinfo {volume} {4}} (\bibinfo {year} {2016})}\BibitemShut
  {NoStop}%
\bibitem [{\citenamefont {Sinitsyn}, \citenamefont {Hengartner},\ and\
  \citenamefont {Nemenman}(2009)}]{sinitsyn2009adiabatic}%
  \BibitemOpen
  \bibfield  {author} {\bibinfo {author} {\bibfnamefont {N.}~\bibnamefont
  {Sinitsyn}}, \bibinfo {author} {\bibfnamefont {N.}~\bibnamefont
  {Hengartner}}, \ and\ \bibinfo {author} {\bibfnamefont {I.}~\bibnamefont
  {Nemenman}},\ }\bibfield  {title} {\enquote {\bibinfo {title} {Adiabatic
  coarse-graining and simulations of stochastic biochemical networks},}\
  }\href@noop {} {\bibfield  {journal} {\bibinfo  {journal} {Proceedings of the
  National Academy of Sciences}\ }\textbf {\bibinfo {volume} {106}},\ \bibinfo
  {pages} {10546--10551} (\bibinfo {year} {2009})}\BibitemShut {NoStop}%
\bibitem [{\citenamefont {Ramaswamy}\ and\ \citenamefont
  {Sbalzarini}(2011)}]{ramaswamy2011partial}%
  \BibitemOpen
  \bibfield  {author} {\bibinfo {author} {\bibfnamefont {R.}~\bibnamefont
  {Ramaswamy}}\ and\ \bibinfo {author} {\bibfnamefont {I.~F.}\ \bibnamefont
  {Sbalzarini}},\ }\bibfield  {title} {\enquote {\bibinfo {title} {A
  partial-propensity formulation of the stochastic simulation algorithm for
  chemical reaction networks with delays},}\ }\href@noop {} {\bibfield
  {journal} {\bibinfo  {journal} {The Journal of chemical physics}\ }\textbf
  {\bibinfo {volume} {134}},\ \bibinfo {pages} {014106} (\bibinfo {year}
  {2011})}\BibitemShut {NoStop}%
\bibitem [{\citenamefont {Jia}\ and\ \citenamefont
  {Kulkarni}(2011)}]{jia2011intrinsic}%
  \BibitemOpen
  \bibfield  {author} {\bibinfo {author} {\bibfnamefont {T.}~\bibnamefont
  {Jia}}\ and\ \bibinfo {author} {\bibfnamefont {R.~V.}\ \bibnamefont
  {Kulkarni}},\ }\bibfield  {title} {\enquote {\bibinfo {title} {Intrinsic
  noise in stochastic models of gene expression with molecular memory and
  bursting},}\ }\href@noop {} {\bibfield  {journal} {\bibinfo  {journal}
  {Physical review letters}\ }\textbf {\bibinfo {volume} {106}},\ \bibinfo
  {pages} {058102} (\bibinfo {year} {2011})}\BibitemShut {NoStop}%
\bibitem [{\citenamefont {Qiu}, \citenamefont {Jia}\ \emph
  {et~al.}(2019)\citenamefont {Qiu}, \citenamefont {Jia} \emph
  {et~al.}}]{qiu2019quantifying}%
  \BibitemOpen
  \bibfield  {author} {\bibinfo {author} {\bibfnamefont {S.}~\bibnamefont
  {Qiu}}, \bibinfo {author} {\bibfnamefont {T.}~\bibnamefont {Jia}},  \emph
  {et~al.},\ }\bibfield  {title} {\enquote {\bibinfo {title} {Quantifying the
  noise in bursty gene expression under regulation by small rnas},}\
  }\href@noop {} {\bibfield  {journal} {\bibinfo  {journal} {International
  Journal of Modern Physics C (IJMPC)}\ }\textbf {\bibinfo {volume} {30}},\
  \bibinfo {pages} {1--14} (\bibinfo {year} {2019})}\BibitemShut {NoStop}%
\bibitem [{\citenamefont {Kumar}\ \emph {et~al.}(2016)\citenamefont {Kumar},
  \citenamefont {Jia}, \citenamefont {Zarringhalam},\ and\ \citenamefont
  {Kulkarni}}]{kumar2016frequency}%
  \BibitemOpen
  \bibfield  {author} {\bibinfo {author} {\bibfnamefont {N.}~\bibnamefont
  {Kumar}}, \bibinfo {author} {\bibfnamefont {T.}~\bibnamefont {Jia}}, \bibinfo
  {author} {\bibfnamefont {K.}~\bibnamefont {Zarringhalam}}, \ and\ \bibinfo
  {author} {\bibfnamefont {R.~V.}\ \bibnamefont {Kulkarni}},\ }\bibfield
  {title} {\enquote {\bibinfo {title} {Frequency modulation of stochastic gene
  expression bursts by strongly interacting small rnas},}\ }\href@noop {}
  {\bibfield  {journal} {\bibinfo  {journal} {Physical Review E}\ }\textbf
  {\bibinfo {volume} {94}},\ \bibinfo {pages} {042419} (\bibinfo {year}
  {2016})}\BibitemShut {NoStop}%
\bibitem [{\citenamefont {Fennell}, \citenamefont {Melnik},\ and\ \citenamefont
  {Gleeson}(2016)}]{fennell2016limitations}%
  \BibitemOpen
  \bibfield  {author} {\bibinfo {author} {\bibfnamefont {P.~G.}\ \bibnamefont
  {Fennell}}, \bibinfo {author} {\bibfnamefont {S.}~\bibnamefont {Melnik}}, \
  and\ \bibinfo {author} {\bibfnamefont {J.~P.}\ \bibnamefont {Gleeson}},\
  }\bibfield  {title} {\enquote {\bibinfo {title} {Limitations of discrete-time
  approaches to continuous-time contagion dynamics},}\ }\href@noop {}
  {\bibfield  {journal} {\bibinfo  {journal} {Physical Review E}\ }\textbf
  {\bibinfo {volume} {94}},\ \bibinfo {pages} {052125} (\bibinfo {year}
  {2016})}\BibitemShut {NoStop}%
\bibitem [{\citenamefont {Lipowski}\ and\ \citenamefont
  {Lipowska}(2012)}]{lipowski2012roulette}%
  \BibitemOpen
  \bibfield  {author} {\bibinfo {author} {\bibfnamefont {A.}~\bibnamefont
  {Lipowski}}\ and\ \bibinfo {author} {\bibfnamefont {D.}~\bibnamefont
  {Lipowska}},\ }\bibfield  {title} {\enquote {\bibinfo {title} {Roulette-wheel
  selection via stochastic acceptance},}\ }\href@noop {} {\bibfield  {journal}
  {\bibinfo  {journal} {Physica A: Statistical Mechanics and its Applications}\
  }\textbf {\bibinfo {volume} {391}},\ \bibinfo {pages} {2193--2196} (\bibinfo
  {year} {2012})}\BibitemShut {NoStop}%
\bibitem [{\citenamefont {Altizer}\ \emph {et~al.}(2006)\citenamefont
  {Altizer}, \citenamefont {Dobson}, \citenamefont {Hosseini}, \citenamefont
  {Hudson}, \citenamefont {Pascual},\ and\ \citenamefont
  {Rohani}}]{altizer2006seasonality}%
  \BibitemOpen
  \bibfield  {author} {\bibinfo {author} {\bibfnamefont {S.}~\bibnamefont
  {Altizer}}, \bibinfo {author} {\bibfnamefont {A.}~\bibnamefont {Dobson}},
  \bibinfo {author} {\bibfnamefont {P.}~\bibnamefont {Hosseini}}, \bibinfo
  {author} {\bibfnamefont {P.}~\bibnamefont {Hudson}}, \bibinfo {author}
  {\bibfnamefont {M.}~\bibnamefont {Pascual}}, \ and\ \bibinfo {author}
  {\bibfnamefont {P.}~\bibnamefont {Rohani}},\ }\bibfield  {title} {\enquote
  {\bibinfo {title} {Seasonality and the dynamics of infectious diseases},}\
  }\href@noop {} {\bibfield  {journal} {\bibinfo  {journal} {Ecology letters}\
  }\textbf {\bibinfo {volume} {9}},\ \bibinfo {pages} {467--484} (\bibinfo
  {year} {2006})}\BibitemShut {NoStop}%
\bibitem [{\citenamefont {Freeman}\ \emph {et~al.}(2006)\citenamefont
  {Freeman}, \citenamefont {Weiss}, \citenamefont {Glynn}, \citenamefont
  {Cross}, \citenamefont {Whitworth},\ and\ \citenamefont
  {Hayes}}]{freeman2006herpes}%
  \BibitemOpen
  \bibfield  {author} {\bibinfo {author} {\bibfnamefont {E.~E.}\ \bibnamefont
  {Freeman}}, \bibinfo {author} {\bibfnamefont {H.~A.}\ \bibnamefont {Weiss}},
  \bibinfo {author} {\bibfnamefont {J.~R.}\ \bibnamefont {Glynn}}, \bibinfo
  {author} {\bibfnamefont {P.~L.}\ \bibnamefont {Cross}}, \bibinfo {author}
  {\bibfnamefont {J.~A.}\ \bibnamefont {Whitworth}}, \ and\ \bibinfo {author}
  {\bibfnamefont {R.~J.}\ \bibnamefont {Hayes}},\ }\bibfield  {title} {\enquote
  {\bibinfo {title} {Herpes simplex virus 2 infection increases hiv acquisition
  in men and women: systematic review and meta-analysis of longitudinal
  studies},}\ }\href@noop {} {\bibfield  {journal} {\bibinfo  {journal} {Aids}\
  }\textbf {\bibinfo {volume} {20}},\ \bibinfo {pages} {73--83} (\bibinfo
  {year} {2006})}\BibitemShut {NoStop}%
\bibitem [{\citenamefont {Jankowski}\ \emph {et~al.}(2018)\citenamefont
  {Jankowski}, \citenamefont {Szymanski}, \citenamefont {Kazienko},
  \citenamefont {Michalski},\ and\ \citenamefont
  {Br{\'o}dka}}]{jankowski2018probing}%
  \BibitemOpen
  \bibfield  {author} {\bibinfo {author} {\bibfnamefont {J.}~\bibnamefont
  {Jankowski}}, \bibinfo {author} {\bibfnamefont {B.~K.}\ \bibnamefont
  {Szymanski}}, \bibinfo {author} {\bibfnamefont {P.}~\bibnamefont {Kazienko}},
  \bibinfo {author} {\bibfnamefont {R.}~\bibnamefont {Michalski}}, \ and\
  \bibinfo {author} {\bibfnamefont {P.}~\bibnamefont {Br{\'o}dka}},\ }\bibfield
   {title} {\enquote {\bibinfo {title} {Probing limits of information spread
  with sequential seeding},}\ }\href@noop {} {\bibfield  {journal} {\bibinfo
  {journal} {Scientific reports}\ }\textbf {\bibinfo {volume} {8}},\ \bibinfo
  {pages} {1--9} (\bibinfo {year} {2018})}\BibitemShut {NoStop}%
\bibitem [{\citenamefont {Jankowski}\ \emph {et~al.}(2017)\citenamefont
  {Jankowski}, \citenamefont {Br{\'o}dka}, \citenamefont {Kazienko},
  \citenamefont {Szymanski}, \citenamefont {Michalski},\ and\ \citenamefont
  {Kajdanowicz}}]{jankowski2017balancing}%
  \BibitemOpen
  \bibfield  {author} {\bibinfo {author} {\bibfnamefont {J.}~\bibnamefont
  {Jankowski}}, \bibinfo {author} {\bibfnamefont {P.}~\bibnamefont
  {Br{\'o}dka}}, \bibinfo {author} {\bibfnamefont {P.}~\bibnamefont
  {Kazienko}}, \bibinfo {author} {\bibfnamefont {B.~K.}\ \bibnamefont
  {Szymanski}}, \bibinfo {author} {\bibfnamefont {R.}~\bibnamefont
  {Michalski}}, \ and\ \bibinfo {author} {\bibfnamefont {T.}~\bibnamefont
  {Kajdanowicz}},\ }\bibfield  {title} {\enquote {\bibinfo {title} {Balancing
  speed and coverage by sequential seeding in complex networks},}\ }\href@noop
  {} {\bibfield  {journal} {\bibinfo  {journal} {Scientific reports}\ }\textbf
  {\bibinfo {volume} {7}},\ \bibinfo {pages} {1--11} (\bibinfo {year}
  {2017})}\BibitemShut {NoStop}%
\bibitem [{\citenamefont {Liu}\ \emph {et~al.}(2018{\natexlab{b}})\citenamefont
  {Liu}, \citenamefont {Zhong}, \citenamefont {Wang}, \citenamefont {Zhou},\
  and\ \citenamefont {Eugene~Stanley}}]{liu2018interactive}%
  \BibitemOpen
  \bibfield  {author} {\bibinfo {author} {\bibfnamefont {Q.-H.}\ \bibnamefont
  {Liu}}, \bibinfo {author} {\bibfnamefont {L.-F.}\ \bibnamefont {Zhong}},
  \bibinfo {author} {\bibfnamefont {W.}~\bibnamefont {Wang}}, \bibinfo {author}
  {\bibfnamefont {T.}~\bibnamefont {Zhou}}, \ and\ \bibinfo {author}
  {\bibfnamefont {H.}~\bibnamefont {Eugene~Stanley}},\ }\bibfield  {title}
  {\enquote {\bibinfo {title} {Interactive social contagions and co-infections
  on complex networks},}\ }\href@noop {} {\bibfield  {journal} {\bibinfo
  {journal} {Chaos: An Interdisciplinary Journal of Nonlinear Science}\
  }\textbf {\bibinfo {volume} {28}},\ \bibinfo {pages} {013120} (\bibinfo
  {year} {2018}{\natexlab{b}})}\BibitemShut {NoStop}%
\bibitem [{\citenamefont {Wang}, \citenamefont {Lan},\ and\ \citenamefont
  {Xiao}(2019)}]{wang2019anomalous}%
  \BibitemOpen
  \bibfield  {author} {\bibinfo {author} {\bibfnamefont {X.}~\bibnamefont
  {Wang}}, \bibinfo {author} {\bibfnamefont {Y.}~\bibnamefont {Lan}}, \ and\
  \bibinfo {author} {\bibfnamefont {J.}~\bibnamefont {Xiao}},\ }\bibfield
  {title} {\enquote {\bibinfo {title} {Anomalous structure and dynamics in news
  diffusion among heterogeneous individuals},}\ }\href@noop {} {\bibfield
  {journal} {\bibinfo  {journal} {Nature human behaviour}\ }\textbf {\bibinfo
  {volume} {3}},\ \bibinfo {pages} {709--718} (\bibinfo {year}
  {2019})}\BibitemShut {NoStop}%
\bibitem [{\citenamefont {Wu}\ \emph {et~al.}(2018)\citenamefont {Wu},
  \citenamefont {Zheng}, \citenamefont {Zhang}, \citenamefont {Wang},
  \citenamefont {Gu},\ and\ \citenamefont {Liu}}]{wu2018model}%
  \BibitemOpen
  \bibfield  {author} {\bibinfo {author} {\bibfnamefont {J.}~\bibnamefont
  {Wu}}, \bibinfo {author} {\bibfnamefont {M.}~\bibnamefont {Zheng}}, \bibinfo
  {author} {\bibfnamefont {Z.-K.}\ \bibnamefont {Zhang}}, \bibinfo {author}
  {\bibfnamefont {W.}~\bibnamefont {Wang}}, \bibinfo {author} {\bibfnamefont
  {C.}~\bibnamefont {Gu}}, \ and\ \bibinfo {author} {\bibfnamefont
  {Z.}~\bibnamefont {Liu}},\ }\bibfield  {title} {\enquote {\bibinfo {title} {A
  model of spreading of sudden events on social networks},}\ }\href@noop {}
  {\bibfield  {journal} {\bibinfo  {journal} {Chaos: An Interdisciplinary
  Journal of Nonlinear Science}\ }\textbf {\bibinfo {volume} {28}},\ \bibinfo
  {pages} {033113} (\bibinfo {year} {2018})}\BibitemShut {NoStop}%
\bibitem [{\citenamefont {Hu}\ \emph {et~al.}(2018)\citenamefont {Hu},
  \citenamefont {Ji}, \citenamefont {Jin}, \citenamefont {Feng}, \citenamefont
  {Stanley},\ and\ \citenamefont {Havlin}}]{hu2018local}%
  \BibitemOpen
  \bibfield  {author} {\bibinfo {author} {\bibfnamefont {Y.}~\bibnamefont
  {Hu}}, \bibinfo {author} {\bibfnamefont {S.}~\bibnamefont {Ji}}, \bibinfo
  {author} {\bibfnamefont {Y.}~\bibnamefont {Jin}}, \bibinfo {author}
  {\bibfnamefont {L.}~\bibnamefont {Feng}}, \bibinfo {author} {\bibfnamefont
  {H.~E.}\ \bibnamefont {Stanley}}, \ and\ \bibinfo {author} {\bibfnamefont
  {S.}~\bibnamefont {Havlin}},\ }\bibfield  {title} {\enquote {\bibinfo {title}
  {Local structure can identify and quantify influential global spreaders in
  large scale social networks},}\ }\href@noop {} {\bibfield  {journal}
  {\bibinfo  {journal} {Proceedings of the National Academy of Sciences}\
  }\textbf {\bibinfo {volume} {115}},\ \bibinfo {pages} {7468--7472} (\bibinfo
  {year} {2018})}\BibitemShut {NoStop}%
\bibitem [{\citenamefont {Miller}(2016)}]{miller2016equivalence}%
  \BibitemOpen
  \bibfield  {author} {\bibinfo {author} {\bibfnamefont {J.~C.}\ \bibnamefont
  {Miller}},\ }\bibfield  {title} {\enquote {\bibinfo {title} {Equivalence of
  several generalized percolation models on networks},}\ }\href@noop {}
  {\bibfield  {journal} {\bibinfo  {journal} {Physical Review E}\ }\textbf
  {\bibinfo {volume} {94}},\ \bibinfo {pages} {032313} (\bibinfo {year}
  {2016})}\BibitemShut {NoStop}%
\end{thebibliography}

%

\end{document}